\documentclass[11pt]{article} 
\usepackage[utf8]{inputenc}  
\usepackage{xcolor} 
\usepackage[margin=1in]{geometry} 
\usepackage{mathtools}        
\usepackage{amsmath}          
\usepackage{amsthm}           
\usepackage{amsfonts}         
\usepackage{nicefrac}         
\usepackage{soul}             
\usepackage{paralist}
\usepackage{siunitx}          

\usepackage{lineno}

\newcolumntype{E}{S[table-format=2.2e-1, table-figures-uncertainty=1]}
\newcolumntype{P}{S[table-format=3.0]}

\usepackage{upgreek}          
\usepackage{dsfont}           

\usepackage[numbers]{natbib} 
\usepackage{graphicx}        
\usepackage{subfig}           
\usepackage{multirow}         
\usepackage[linesnumbered,ruled,vlined]{algorithm2e}
\SetKwInput{KwInput}{Input}
\SetKwInput{KwOutput}{Output}
\SetKwBlock{Begin}{Begin}{end}
\SetKwFor{ForEachPar}{for each (in parallel)}{do}{end}
\usepackage{comment}          
 
\usepackage{authblk}         
 
\usepackage{hyperref}         
\usepackage{indentfirst}      
\usepackage{bm}        
\usepackage{mathrsfs}  
\usepackage{array,tabularx,booktabs}

\newcolumntype{L}[1]{>{\raggedright\arraybackslash}p{#1}}
\newcolumntype{C}[1]{>{\centering\arraybackslash}p{#1}}  
\newcolumntype{R}[1]{>{\raggedleft\arraybackslash}p{#1}}  

\graphicspath{ {Figures/} }

\title{Ensemble-Based Data Assimilation for Material Model Characterization in High-Velocity Impact}
\author[1]{Rong Jin}
\author[2,3]{Guangyao Wang}
\author[1]{Xingsheng Sun\footnote{Corresponding author: xingsheng.sun@uky.edu}}

\affil[1]{Department of Mechanical and Aerospace Engineering, University of Kentucky, Lexington, KY 40506, USA}
\affil[2]{Centre for Regional Oceans \& Department of Ocean Science and Technology, Faculty of Science and Technology, University of Macau, 999078, Macau}
\affil[3]{State Key Laboratory of Internet of Things for Smart City, University of Macau, 999078, Macau}
\date{}
\begin{document}

\maketitle
\begin{abstract}

High-fidelity simulations are essential for understanding and predicting the behavior of materials under high-velocity impact (HVI) in both fundamental research and practical applications. However, their accuracy relies on material models and parameters that are traditionally obtained through manual fitting to multiple time- and labor-intensive experiments. In this study, we develop an ensemble-based data assimilation (DA) framework that automatically and simultaneously calibrates selected plasticity, fracture, and equation of state (EOS) model parameters in HVI simulations using measured data from a single HVI test. The framework integrates Smoothed Particle Hydrodynamics for HVI simulations, the ensemble Kalman filter (EnKF) for parameter refinement, and adaptive covariance inflation to mitigate underestimation of uncertainty. We first show that, for a simple test problem, the framework achieves a computational efficiency at least one order of magnitude higher than traditional Markov chain Monte Carlo methods at comparable identification accuracy. We then demonstrate the approach using synthetic back-face deflection data of an AZ31B magnesium plate to identify representative parameters in the Johnson-Cook plasticity and fracture models and the Mie-Gr\"uneisen EOS. Test cases include under- and over-biased initial guesses, as well as limited observational data. The results show that, when the observational data are sufficiently sensitive to the parameters, the EnKF-based framework can accurately and efficiently recover the material parameters in five iterations with convergent ensemble standard deviation. In contrast, insensitive parameters tend to converge to incorrect values and are characterized by persistently large ensemble standard deviations. Furthermore, studies on observation quantity reveal that the limited data can still achieve convergence to the true values but with more iterations. Under extreme prior bias, the proposed parameter rejuvenation strategy effectively ensures that sensitive parameters migrate toward the true values even when the truth lies outside the initial ensemble spread. In practical applications where the true parameters are unknown, the ensemble standard deviation thus provides a diagnostic tool to assess parameter sensitivity, identifiability, and potential non-uniqueness. Overall, this study demonstrates the robustness and efficiency of the proposed DA framework for material model characterization in HVI problems.

\end{abstract}

\section{Introduction}

High-velocity impact (HVI) occurs when a projectile strikes a target at speeds ranging from a few hundred meters per second to several kilometers per second, generating extreme stresses, strain rates, and temperatures over microsecond to millisecond timescales~\cite{zukas1990high}. This phenomenon plays a critical role in aerospace engineering, planetary science, and defense applications. Under such conditions, stresses often exceed the material’s yield strength, causing the response to become increasingly ``fluid-like'' while still influenced by material strength. HVI is inherently a multi-physics, multi-scale problem~\cite{islam2025ionization, liu2021hierarchical, braroo2025uncertainty}, coupling elastic-plastic deformation, material failure, shock wave propagation, and thermodynamic processes over a broad spectrum of strain rates and temperatures.

Understanding material behavior under HVI conditions is essential for designing protective systems, optimizing materials, and ensuring structural integrity. The impact results in stress waves or shock waves that may initially be elastic but quickly transition to plastic or even hydrodynamic character as the material yields. As the impact velocity increases, the induced pressures may eventually greatly exceed the material’s flow stress; the response then effectively enters a hydrodynamic regime where density and inertia dominate over strength~\cite{signetti2022transition, signetti2021characterization}. In this regime, strong shock waves propagate through the material, producing abrupt changes in density, temperature, and entropy. HVI events can give rise to complex phenomena such as impact flash~\cite{simpson2023first} and cratering \cite{stradling1993ultra}, and many material failure mechanisms, including plugging~\cite{klepaczko1998remarks}, petaling~\cite{wierzbicki1999petalling}, spalling~\cite{meyers1983dynamic}, and fragmentation~\cite{kipp1993numerical}. Throughout the impact process, a portion of the projectile’s kinetic energy is converted into internal energy, increasing the temperature and stored strain energy of both the projectile and the target, while another portion is transferred to the kinetic energy of the target motion~\cite{signetti2022transition, jin2024characterization}. At extreme velocities, additional dissipation mechanisms can occur, including phase transformations~\cite{zhu2016phase} and plasma generation~\cite{islam2023fluid}.

Because of the extreme complexity of HVI, including strong non-linearity, singularities, strong dependence on impact configurations, and sensitivity to multiple physical processes, closed-form analytical solutions are generally not available. Consequently, numerical simulations have become the primary tool for investigating such events, employing methods such as Lagrangian finite elements (e.g., LS-DYNA, Abaqus/Explicit), Eulerian formulations (e.g., CTH~\cite{mcglaun1990cth}, M2C~\cite{islam2025ionization}), Arbitrary Lagrangian-Eulerian (ALE) approaches~\cite{nazem2008arbitrary}, and mesh-free techniques such as Smooth Particle Hydrodynamics (SPH)~\cite{hayhurst1997cylindrically} and peridynamics~\cite{silling2004peridynamic}. Within these simulations, three core components govern material modeling and, therefore, predictive accuracy. First, constitutive models---such as Johnson-Cook (JC)~\cite{johnson1983constitutive}, Zerilli-Armstrong~\cite{zerilli1987dislocation}, or physically based dislocation models~\cite{meyers2002constitutive} for metals---define the material's stress-strain response under extreme strain rates and temperatures, capturing elastic-plastic behavior, strain-rate sensitivity, and thermal softening. Second, material failure and fracture models, such as Johnson-Cook damage~\cite{johnson1985fracture} and cohesive zone~\cite{elices2002cohesive} formulations, describe the initiation and evolution of damage, enabling the simulation of failure modes such as plugging, spalling, petaling, and fragmentation. Third, volumetric response, typically represented by an equation of state (EOS) such as Mie-Gr\"uneisen~\cite{gruneisen1912theorie} or Tillotson~\cite{tillotson1962metallic}, governs material compressibility and shock behavior, accounting for density changes, phase transitions, and thermodynamic effects. 

Calibrating material parameters in these models presents significant challenges. They are often estimated by manually tuning them until a visually acceptable agreement is achieved with targeted experimental results. For instance, for constitutive models spanning multiple orders of strain rates~\cite{eswarappa2024high}, parameters at quasi-static rates can be determined using strain-stress curves measured from conventional tensile testing machines. In contrast, strain-rate-dependent parameters require more specialized, time- and labor-intensive techniques, such as instrumented drop tower apparatus~\cite{lee2004direct} (for strain rates on the order of $10$-$10^{2} \ \mathrm{s}^{-1}$) and Kolsky bar experiments~\cite{kolsky1949investigation} (for strain rates on the order of $10^{3} \ \mathrm{s}^{-1}$). As the number of material models increases, so does the number of targeted experiments required. Likewise, as the number of parameters grows, the volume of experimental data needed to sufficiently explore the parameter space increases significantly. Coupled with the high computational cost of modern high-fidelity simulations, this makes parameter optimization both time-consuming and expensive. Furthermore, the risk of non-unique or over-fitted solutions increases when parameter dimensionality is high and experimental datasets are limited or noisy. Traditional optimization approaches, which minimize predefined error norms, typically return a single best-fit parameter set but do not provide a probabilistic description of the parameter space, thereby limiting the ability to quantify uncertainty in model inputs and predictions.

Bayesian model calibration provides a rigorous statistical framework for aligning computationally intensive physics models with experimental data to estimate the input parameters that yield the best agreement~\cite{kennedy2001bayesian, higdon2008computer}. In addition to producing point estimates, this approach quantifies parameter uncertainty through probability distributions and can explicitly account for systematic discrepancies between the computational model and experimental measurements, incorporating both their estimates and uncertainties. Traditional derivative-free Bayesian calibration methods, such as Markov chain Monte Carlo (MCMC)~\cite{walters2018bayesian, nguyen2021bayesian}, sample from the posterior distribution by generating a large number of iterations---often exceeding $10^{4}$---to achieve statistical convergence. However, when each forward simulation is computationally expensive, the required number of runs makes MCMC prohibitively costly, limiting its practicality for real-world HVI calibrations. To address this challenge, surrogate models such as Gaussian process regression~\cite{higdon2008computer} are often employed to approximate the high-fidelity model, significantly reducing computational cost while retaining acceptable predictive accuracy. However, surrogate models introduce an additional layer of approximation error, and their predictive reliability may deteriorate when extrapolating beyond the range of the training data, especially for highly nonlinear or discontinuous HVI phenomena.

In this study, we have developed a computational framework based on the ensemble Kalman filter (EnKF)~\cite{evensen1994sequential, evensen2009data, cao2022bayesian} for the Bayesian calibration of HVI simulations, with the goal of automatically and simultaneously calibrating multiple material models using data from a single HVI test. The EnKF is a Monte Carlo-based approach that represents the system state using an ensemble of realizations. It operates in two main steps: (1) prediction, where the ensemble is propagated forward in time using the forward model; and (2) analysis, where the predicted state and its uncertainty are updated to incorporate observational data. In the analysis step, the EnKF computes the sample covariance directly from the ensemble, replacing the explicit covariance matrix and forecast operator used in the traditional Kalman filter, thereby reducing computational cost for nonlinear problems~\cite{iglesias2013ensemble, katzfuss2016understanding}. Due to its non-intrusive nature, the EnKF has been widely applied for parameter estimation in problems where the forward model is computationally expensive and provided as a black box, making differentiation impractical. Applications of the EnKF in model calibration include viscoelastic material calibration using bubble-collapse observations~\cite{spratt2021characterizing}, ocean wave forecast model calibration using radar measurements~\cite{wang2021phase, wang2022phase}, multi-phase-field model calibration with synthetic microstructure data~\cite{yamanaka2019ensemble}, and phase-field model calibration using twin-experiment data~\cite{sasaki2018data}, among others.

We apply the EnKF framework to a case study involving the HVI of an AZ31B magnesium alloy plate impacted by a  steel sphere~\cite{malhotra2025modified}, simulated using a high-fidelity SPH numerical model. The model to be calibrated incorporates the JC plasticity and fracture models together with the Mie-Gr\"uneisen EOS for the target plate. A key element of the framework is the selection of the plate's back-face deflection---which can be measured via high-speed digital gradient sensing (DGS) ~\cite{malhotra2025modified}---as the observable for data assimilation. This metric is advantageous due to its: (1) full-field capability, providing comprehensive displacement data over the entire back-face of the target; (2) time-resolved nature, capturing the transient dynamics of momentum transfer, stress-wave propagation, and damage evolution in each frame; and (3) direct measurability, eliminating the need for additional transducers or complex inverse image reconstructions beyond standard DIC techniques. In this study, we define an artificial time step that is distinct from the physical time step within the traditional EnKF formulation for time-dependent problems. Each artificial time step corresponds to a Kalman filtering iteration, where the observation consists of a time series of back-face deflection data from a complete HVI test. The proposed EnKF framework is validated using synthetic datasets designed to investigate the influence of the initial guess mean, as well as the impact of limited observational data.

The remainder of the paper is organized as follows. Section~\ref{sec:problem} formulates the HVI model-calibration problem. Section~\ref{sec:method} outlines the fundamentals of the SPH numerical method and presents the governing equations for material modeling. Section~\ref{sec:DA} introduces the EnKF equations and describes the development of the EnKF framework. Section~\ref{sec:comp} compares this framework with a traditional MCMC method using a ring-down problem. Section~\ref{sec:nexperiments} details the numerical experiments, including the specification of model inputs and outputs, sensitivity screening, generation of synthetic observations, and presentation of calibration results. Section~\ref{sec:discussion} discusses the main findings, and Section~\ref{sec:summary} concludes the paper with a concise summary.

\section{Problem Statement}
\label{sec:problem}

The HVI model-calibration problem can be formulated as a general inverse problem: given observational data $\bm{y} \in \mathbb{R}^{N_{y}}$, infer the unknown system parameters $\bm{u}\in \mathbb{R}^{N_{u}}$ through a forward model $\mathcal{G}$. We adopt the standard observation model 
\begin{equation}
\label{eq:inverse_form}
    \bm{y}=\mathcal{G}(\bm{u})+\boldsymbol{\eta},
    \quad
    \boldsymbol{\eta} \sim \mathcal{N}(\mathbf{0},\bf{R}),
\end{equation}
where the deterministic forward operator $\mathcal{G}:\mathbb{R}^{N_u} \rightarrow \mathbb{R}^{N_y}$ maps the unknown parameters to the observational space, and $\boldsymbol{\eta} \in \mathbb{R}^{N_{y}}$ represents the measurement noise, modeled as zero-mean Gaussian with known covariance $\bf{R}$. In this study, $\mathcal{G}$ is realized by a complete SPH simulation, while $\bm{u}$ collects the material model parameters to be calibrated. The observational data $\bm{y}$ consist of time-series back-face deflection measurements, assumed to be obtained from a single HVI test. We further assume $\mathcal{G}$ is perfect, so all discrepancy arises from imperfect parameters; that is, if $\bm{u}$ matches the true values, then $\mathcal{G}(\bm{u})$ coincides with the true system response. The calibration task is therefore to identify $\bm{u}$ that best matches the noisy observations $\bm{y}$.

\section{Forward Simulation and Material Modeling}
\label{sec:method}

For completeness and ease of reference, this section briefly summarizes the governing equations of the SPH method and the material models employed in the HVI simulations, which together define the forward operator $\mathcal{G}$. A comprehensive review of the SPH method can be found in Ref.~\cite{monaghan2012smoothed}, while a review of material models applied to HVI problems is provided in Ref.~\cite{signetti2022transition}.

\subsection{Smooth Particle Hydrodynamics}\label{subsec:SPH}  

SPH is a mesh-free, Lagrangian method for solving fluid and solid mechanics problems~\cite{monaghan1992smoothed, gingold1977smoothed}. In SPH, the continuum is represented by a set of particles that carry physical quantities such as mass, velocity, and energy. Field variables and their derivatives are approximated by kernel-weighted averages over neighboring particles, which avoids the need for a computational grid. This makes SPH particularly effective for problems involving large deformations, free surfaces, fracture, and impact, such as HVI simulations~\cite{johnson1996sph}. Specifically, SPH approximates a continuous field $f$ at position $\bm{r}$ by the kernel function $W$~\cite{koschier2020smoothed}
\begin{equation}
\label{eq:SPHcontinuusapprox}
    f(\bm{r}) \simeq (f*W) (\bm{r}) = \int f(\bm{r}') W(\bm{r}-\bm{r}',h)\, \text{d}\bm{v}',
\end{equation}
where $\text{d}\boldsymbol{v}'$ denotes the integration variable over the volume associated with the position $\bm{r}'$, $h$ represents the smoothing length, and $W(\bm{r}-\bm{r}',h)$ is the kernel evaluated at position $\bm{r}-\bm{r}'$. 
Introducing $\text{d} m^{\prime} = \rho(\bm{r}^{\prime}) \text{d} v^{\prime}$, where $\rho$ denotes the density, yields the particle approximation~\cite{koschier2020smoothed,fraga2019smoothed}
\begin{equation}
    \int \frac{f(\bm{r}')}{\rho(\bm{r}')} W(\bm{r}-\bm{r}',h) \underbrace{\rho\left(\bm{r}^{\prime}\right) \text{d} v^{\prime}}_{\text{d} m^{\prime}} \approx \sum_{b \in \mathcal{F}_a} f_b \frac{m_b }{\rho_b} W_{ab},
\end{equation}
where $a$ is the particle at position $\bm{r}$, $\mathcal{F}_a$ the neighbor set of particle $a$, $m_b$ the particle mass, and $\rho_b$ the local density of particle $b$. For simplicity, the abbreviations $f_b:=f(\bm{r}_b)$ and $W(\bm{r}_a - \bm{r}_b,h) := W_{ab}$ are used. 

For three-dimentional problems, the cubic spline kernel is adopted~\cite{randles1996smoothed}
\begin{equation}\label{eq:cubicspline}
    W(\bm{r}, h) = \frac{1}{\pi h^3} \begin{cases}
        1 - \frac{3}{2}\xi^2 + \frac{3}{4}\xi^3, & 0 \leq \xi < 1 \\
        \frac{1}{4}(2 - \xi)^3, & 1 \leq \xi < 2 \\
        0, & \xi \geq 2
    \end{cases}
\end{equation}
where $\xi = \frac{\lVert \bm{r} \rVert}{h}$, with $\lVert \bm{r} \rVert$ being the particle distance. The kernel is normalized such that its integral over the entire domain equals unity.

The governing equations of mass, momentum, and energy conservation can be discretized using the SPH method~\cite{fraga2019smoothed} in conjunction with a symmetric formulation: 
\begin{equation}\label{eq:masssph}
    \dot{\rho}_a = \rho_a \left[ \sum_{b \in \mathcal{F}_a} \frac{m_b}{\rho_b} (\bm{v}_a - \bm{v}_b) \right] \cdot \nabla_a W_{ab},
\end{equation}
\begin{equation}\label{eq:momentumsph}
    \dot{v}_a^{i} = \sum_{b \in \mathcal{F}_a} m_b \left[ \frac{\sigma^{ij}_a}{\rho_a} + \frac{\sigma^{ij}_b}{\rho_b} \right] \nabla_a^{j} W_{ab},
\end{equation}
\begin{equation}
    \dot{e}_a = \frac{1}{2} \sum_{b \in \mathcal{F}_a} m_b \left[ \frac{\sigma^{ij}_a}{\rho^2_a} + \frac{\sigma^{ij}_b}{\rho^2_b} \right] \left[v_{a}^{j} - v_{b}^{j}\right] \nabla_a^{j} W_{ab}.
\end{equation}
Here, the reference particle $a$ has density $\rho_a$, velocity $\bm{v}_a$, Cauchy stress tensor $\boldsymbol{\sigma}_a$, and specific internal energy $e_a$. The superscripts $i$ and $j$ denote vector and tensor indices in the Cartesian coordinate system. The gradient operator $\nabla_a W_{ab}$ represents the kernel derivative with respect to the position of particle $a$, while keeping particle $b$ fixed. In practical implementations, weak SPH formulations of these conservation laws are employed, together with standard corrective terms, to ensure numerical stability and consistency~\cite{lsdynatheory}.

\subsection{Johnson-Cook Plasticity Model}\label{subsubsec:JCpf}
The JC plasticity model~\cite{johnson1983constitutive} is extensively used to characterize the elastic to plastic behavior of ductile metals over broad ranges of strain rates and temperatures, especially under HVI conditions. Within the plasticity model, the flow stress is decomposed into components attributed to strain hardening, strain rate sensitivity, and thermal softening
\begin{equation}
  \sigma_{\rm{y}} = 
  \bigl[A + B\,\epsilon_{\rm{pl}}^{\,n}\bigr]
  \bigl[1 + C\ln\dot{\epsilon}_{\rm{pl}}^\ast\bigr]
  \bigl[1 - {T^\ast}^{m}\bigr],
\label{eq:JCplasticity}
\end{equation}
where $\sigma_{\rm{y}}$ is the true flow stress, and $\epsilon_{\rm{pl}}$ is the equivalent plastic strain. The normalized plastic strain rate $\dot{\epsilon}_{\rm{pl}}^\ast$ and normalized temperature $T^\ast$ are
\begin{equation}
  \dot{\epsilon}_{\rm{pl}}^\ast = \frac{\dot{\epsilon}_{\rm{pl}}}{\dot{\epsilon}_{\rm{pl0}}}, 
  \quad
  T^\ast = \frac{T-T_0}{T_{\rm{m}}-T_0},
\label{eq:JCeStarTstar}
\end{equation}
respectively, where $\dot{\epsilon}_{\rm{pl}}$ is the plastic strain rate, $\dot{\epsilon}_{\rm{pl0}}$ is a reference plastic strain rate, $T_0$ is the room temperature, and $T_{\rm{m}}$ is the melting temperature of the material. The material parameters requiring calibration are: $A$, the quasi-static yield stress; $B$, the hardening modulus; $n$, the strain hardening exponent; $C$, the strain-rate sensitivity coefficient; and $m$, the thermal softening exponent. 

\subsection{Johnson-Cook Fracture Model}
We assume that the failure behavior of the target plate can be described by the JC fracture model~\cite{johnson1985fracture}. Specifically, the damage of an SPH particle is characterized by a cumulative damage parameter
\begin{equation}
\label{eq:JCfracture}
  D = \sum \frac{\Delta\epsilon_{\rm{pl}}}{\epsilon_{\rm{pl}}^{\rm{f}}},
\end{equation}
where the summation is conducted over time steps, and $\Delta\epsilon_{\rm{pl}}$ denotes the plastic strain increment at each time step. The fracture strain $\epsilon_{\rm{pl}}^{\rm f}$ is defined as
\begin{equation}
\label{eq:JCfractureepsilon}
  \epsilon_{\rm{pl}}^{\rm f} =
  \bigl[D_1 + D_2\exp(D_3\sigma^\ast)\bigr]
  \bigl[1 + D_4\ln\dot{\epsilon}_{\rm{pl}}^\ast\bigr]
  \bigl[1 + D_5 T^\ast\bigr],
\end{equation}
where the stress triaxiality $\sigma^\ast = p/\sigma_{\rm eq}$ is given by the ratio of the hydrostatic pressure $p$ to the von- Mises equivalent stress $\sigma_{\rm eq}$. The normalized plastic strain rate $\dot{\epsilon}_{\rm pl}^\ast$ and normalized temperature $T^\ast$ follow the same definitions as in Eq.~\eqref{eq:JCeStarTstar}. The material constants $D_1$ through $D_5$ are model parameters that must be calibrated to capture fracture initiation and evolution. Within this formulation, fracture occurs once the cumulative damage parameter reaches the critical value $D=1$.

\subsection{Mie-Gr\"uneisen Equation of State}\label{subsubsec:GEOS}

In high-pressure scenarios, such as those arising in HVI, the volumetric response of materials is often described by the Mie-Gr\"uneisen EOS~\cite{gruneisen1912theorie}. This formulation distinguishes between compressed and expanded states~\cite{lsdynaii}. For the compressed state, the pressure is related to density and internal energy as
\begin{equation}\label{eq:GruneisenEOScomp}
  p = 
  \frac{\rho_0 C_{\rm s}^2\,\mu\bigl[1+\tfrac{1}{2}\bigl(1-\gamma_0\bigr)\mu-\tfrac{\alpha}{2}\mu^2\bigr]}
       {\bigl[1-(S_1-1)\mu-S_2\mu^2/(\mu+1)-S_3\mu^3/(\mu+1)^2\bigr]^2}
  +\bigl(\gamma_0+\alpha\mu\bigr)e_0,
\end{equation}
whereas in the expanded state it simplifies to
\begin{equation}\label{eq:GruneisenEOSexp}
  p = \rho_0 C_{\rm s}^2\,\mu + \bigl(\gamma_0+\alpha\mu\bigr)e_0 .
\end{equation}
In this context, $\mu=\rho/\rho_{0}-1$ denotes the nominal volumetric compression, where $\rho$ is the current mass density and $\rho_{0}$ is the reference density. $e_0$ is the specific internal energy per unit mass. $C_{\rm{s}}$ denotes the zero-pressure bulk sound speed. $S_1$, $S_2$, $S_3$, $\gamma_0$, and $\alpha$ are model parameters. Specifically, the relationship between $C_{\rm{s}}$, $S_1$, $S_2$, and $S_3$ can be expressed through the shock-particle velocity relation
\begin{equation}
    v_{\rm{s}} = C_{\rm{s}} + S_1 v_{\rm{p}} + S_2 v_{\rm{p}}^{2} + S_3 v_{\rm{p}}^{3},
\end{equation}
where $v_{\rm{s}}$ is the shock velocity and $v_{\rm{p}}$ is the particle velocity behind the shock. The coefficients $S_1$, $S_2$, and $S_3$ provide successive higher-order corrections to the slope and curvature of the $v_{\rm{s}}$-$v_{\rm{p}}$ curve, thereby enabling the EOS to reproduce the material’s principal Hugoniot over a broad pressure range. In this study, we set $S_2=S_3=0$, which reduces the relation to a linear dependence between shock and particle velocities. As a result, the material parameters requiring calibration in EOS include $S_1$, $\gamma_0$, and $\alpha$. 

\section{Data Assimilation}\label{sec:DA}

Data assimilation provides a powerful framework for addressing the inverse problem formulated in Eq.~(\ref{eq:inverse_form}). It combines observational data with numerical model predictions to yield the best possible estimate of a system’s state. Two primary challenges in selecting an assimilation method are the nonlinear nature of the system and the high computational cost of the forward model. The former rules out the standard linearized Kalman filter~\cite{kalman1960new}, while the latter renders its nonlinear extensions (e.g., extended Kalman filter~\cite{ribeiro2004kalman}) computationally prohibitive. To overcome these limitations, ensemble-based methods~\cite{evensen1994sequential} are employed. These methods balance computational efficiency with the ability to capture nonlinear systems by approximating the state covariance using the statistics of a finite (and typically small) ensemble. In this study, we adopt an EnKF-based data assimilation approach.

In the traditional EnKF framework, the state vector contains all state variables at a given physical time step. When unknown model parameters are to be estimated, they are appended to the state vector and treated as additional, time-varying state variables. In this study, however, we introduce an artificial time step that is distinct from the physical time step used in conventional time-dependent EnKF formulations. Each artificial time step corresponds to one Kalman filtering iteration, in which the observation consists of a full time series of back-face deflection data obtained from a single HVI test. The resulting discrete "dynamical" system can then be expressed as~\cite{carrassi2018data}
\begin{subequations}
\label{eq:DAmodel}
    \begin{alignat}{2}
    \bm{x}_{l+1} &= \mathcal{M}(\bm{x}_l) + \boldsymbol{\omega}_{l+1}, \quad &
    \boldsymbol{\omega}_{l+1} &\sim \mathcal{N}(\mathbf{0},\mathbf{\Sigma}_{\omega}),\\
    \bm{y}_{l+1} &= \mathcal{H}(\bm{x}_{l+1}) + \boldsymbol{\nu}_{l+1}, &
    \boldsymbol{\nu}_{l+1}  &\sim \mathcal{N}(\mathbf{0},\mathbf{\Sigma}_{\nu}),
    \end{alignat}
\end{subequations}
where $l$ denotes the artificial time step. The augmented state vector $\bm{x}_{l} \in \mathbb{R}^{N_y + N_u}$ at time step $l$ contains both the unknown material parameters $\bm{u}$ and the model output $\mathcal{G}(\bm{u})$
\begin{align}
\label{eq:aug_state}
    \bm{x} &= \begin{bmatrix}
        \mathcal{G}(\bm{u}) \\
        \bm{u} 
            \end{bmatrix}.
\end{align}
The observation vector $\bm{y}_l \in \mathbb{R}^{N_y}$ corresponds to the data at artificial time step $l$. The forecast operator $\mathcal{M}$ propagates the state vector from step $l$ to $l+1$. In this formulation, $\mathcal{M}$ maps the augmented state vector $\bm{x}$ to itself since they remain constant in the physical model. Hence, $\mathcal{M}$ is often referred to as the persistence model. The observation operator $\mathcal{H}$ maps the state vector to the observation space. In this study, $\mathcal{H}$ is linear and extracts the first $N_y$ components of $\bm{x}$, i.e.,  
\begin{equation}
    \mathcal{H}(\bm{x}) = \mathbf{H}\bm{x}, \quad 
\mathbf{H} = \bigl[\,\mathbf{I}_{N_y} \;\; \mathbf{0}_{N_y\times N_u}\,\bigr],
\end{equation}
where $\mathbf{I}_{N_y}$ is the $N_y\times N_y$ identity matrix and $\mathbf{0}_{N_y\times N_u}$ is the zero matrix. In Eq.~(\ref{eq:DAmodel}), the random vectors $\boldsymbol{\omega}_l$ and $\boldsymbol{\nu}_l$ represent model and measurement errors, respectively, and are modeled as mutually independent Gaussian random variables with known covariance matrices $\bm{\Sigma}_{\omega}$ and $\bm{\Sigma}_{\nu}$.

\subsection{Ensemble Kalman Filter}
\label{subsec:EnKF}

The EnKF is a sequential data assimilation method that combines model forecasts with observations to estimate the evolving state of a system~\cite{evensen1994sequential}. Uncertainty is represented by an ensemble of $N_{\rm e}$ model realizations, which are propagated forward through the nonlinear dynamics to form a forecast ensemble. When new observations become available, the ensemble is updated via a Kalman-like correction, where the forecast error covariance is approximated from ensemble statistics. 

In this study, we denote by $\bm{x}_l^{(i)}$ the $i$-th ensemble member (augmented state vector) at the $l$-th EnKF iteration. The ensemble mean $\overline{\bm{x}}_l$ and the forecast error covariance matrix $\mathbf{P}_l$ are then computed as  
\begin{equation}
\label{eq:xbar}
    \overline{\bm{x}}_l 
    = \frac{1}{N_{\rm e}} \sum_{i=1}^{N_{\rm e}} \bm{x}_l^{(i)}
    = \frac{1}{N_{\rm e}} \sum_{i=1}^{N_{\rm e}}
    \begin{bmatrix}
        \mathcal{G}\!\big(\bm{u}_l^{(i)}\big) \\
        \bm{u}_l^{(i)}
    \end{bmatrix},
\end{equation}
and  
\begin{equation}
\label{eq:Pforecast}
    \mathbf{P}_l = \frac{1}{N_{\rm e}-1}\sum_{i=1}^{N_{\rm e}} 
    \bigl( \bm{x}^{(i)}_l - \overline{\bm{x}}_l\bigr) 
    \bigl( \bm{x}^{(i)}_l - \overline{\bm{x}}_l\bigr)^\text{T}.
\end{equation}
At each iteration, the ensemble mean $\overline{\bm{x}}_l$ is taken as the best estimate of the system state.

The EnKF starts with initializing the unknown parameters with a guess of the initial condition $\bm{u}_0$ as mean and a given covariance $\mathbf{C}_0$ corresponding to the expected error covariance. Each ensemble member is independently sampled from a Gaussian distribution with mean $\bm{u}_0$ and the assumed covariance matrix $\mathbf{C}_0$ 
\begin{equation}
\label{eq:initial_ensemble}
    \bm{u}^{(i)}_0 \sim \mathcal N(\bm{u}_0,\mathbf{C}_0), 
    \quad 
    i=1,\dots,N_{\rm{e}}.
\end{equation}

The filter then proceeds iteratively with a \textit{forecast step} and an \textit{analysis step}. At iteration $l$, in the forecast step, the model output $\mathcal{G}(\bm{u}_{l}^{(i)})$ is obtained by running the SPH simulation, and the state vector $\bm{x}_{l}^{(i)}$ is formed by augmenting the model output with the parameter vector. Then each state vector is propagated through the forecast operator $\mathcal{M}$, giving $\hat{\bm{x}}_{l+1}^{(i)}=\bm{x}_{l}^{(i)}$.  

In the analysis step, if an experimental observation is available, it is used to correct the forecast $\hat{\bm{x}}_{l+1}^{(i)}$ according to  
\begin{equation}
\label{eq:enkf_update}
    \bm{x}_{l+1}^{(i)} = \hat{\bm{x}}_{l+1}^{(i)} + \mathbf{K}_l 
    \left[ \bm{y}_{\rm{pert}}^{(i)} - \mathbf{H} \hat{\bm{x}}_{l+1}^{(i)}\right], 
    \quad 
    i=1,\dots,N_{\rm{e}},
\end{equation}
where the Kalman gain is given by  
\begin{equation}
\label{eq:enkf_gain}
    \mathbf{K}_l = \mathbf{P}_l \mathbf{H}^\text{T} 
    \left[ \mathbf{H} \mathbf{P}_l \mathbf{H}^\text{T} + \mathbf{R} \right]^{-1}.
\end{equation}
In Eq.~(\ref{eq:enkf_update}), $\bm{y}_{\rm pert}^{(i)}$ denotes the perturbed observation
\begin{equation}
\label{eq:obs}
\bm{y}_{\rm pert}^{(i)} = \bm{y}_{\rm obs} + \boldsymbol{\eta}^{(i)},
\end{equation}
where $\bm{y}_{\rm obs}$ is the observation vector and $\boldsymbol{\eta}^{(i)}$ is a zero-mean Gaussian perturbation, i.e., $\boldsymbol{\eta}^{(i)} \sim \mathcal{N}(\mathbf{0}, \mathbf{R})$. The superscript $(i)$ implies that the observation error is sampled in each ensemble.

After the analysis step, covariance inflation is applied to the ensemble to mitigate the typical underestimation of variance arising from the use of finite (and generally small) ensembles. The procedure then proceeds with the next forecast step. It is worth noting that the gain in Eq.~(\ref{eq:enkf_gain}), derived entirely from the ensemble statistics, balances the observation-forecast mismatch against the forecast and noise covariances without requiring tangent or adjoint operators. This makes the EnKF particularly attractive for black-box, computationally intensive, or non-differentiable forward models~\cite{wang2022phase}.

\subsection{Covariance Inflation}
\label{subsec:CI}

In practical applications of EnKF, the limited ensemble size often leads to an underestimate of the true forecast error covariance. This issue arises because ensemble statistics computed from a small number of realizations tend to underestimate variance and produce false correlations, especially in high-dimensional systems. Such variance underestimation can cause the filter to become overconfident in its predictions, which in turn reduces the weight given to observations and may eventually lead to filter divergence. To mitigate this problem, we employ covariance inflation to artificially increase the ensemble spread, thereby compensating for sampling errors and maintaining a realistic representation of the forecast uncertainty.

Specifically, after the analysis step, we correct each ensemble realization according to  
\begin{equation}
\label{eq:infl_generic}
  \bm{x}_l^{(i)}
  \;\leftarrow\;
  \bar{\bm{x}}_l
  + \boldsymbol{\beta}_l \odot \big(\bm{x}_l^{(i)}-\bar{\bm{x}}_l \big)
  + \bm{\lambda}^{(i)}_l, 
  \quad i=1,\dots,N_{\rm e},
\end{equation}
where $\boldsymbol{\beta}_l \in \mathbb{R}^{N_y+N_u}$ and $\bm{\lambda}^{(i)}_l \in \mathbb{R}^{N_y+N_u}$ denote the multiplicative and additive inflation parameters, respectively, and $\odot$ is the Hadamard product. Following Ref.~\cite{spratt2021characterizing}, we adopt the Relaxation-to-Prior-Spread (RTPS) strategy~\cite{whitaker2012evaluating}, employing zero additive inflation ($\bm{\lambda}^{(i)}_l=\bm{0}$) and a component-wise multiplicative factor  
\begin{equation}
\label{eq:rtps}
    \boldsymbol{\beta}_l{(j)} 
    = 1 + \kappa \,
    \frac{\sqrt{\mathbf{P}_{l-1}(j,j)} - \sqrt{\mathbf{P}_{l}(j,j)}}{\sqrt{\mathbf{P}_{l}(j,j)}},
    \quad j=1,\dots,N_y+N_u,
\end{equation}
where $\boldsymbol{\beta}_l{(j)}$ is the $j$-th component of the multiplicative inflation vector $\boldsymbol{\beta}_l$, and $\kappa$ is a scalar. $\mathbf{P}_l(j,j)$ denotes the $j$-th diagonal entry of the covariance matrix $\mathbf{P}_l$. Thus, $\sqrt{\mathbf{P}_{l-1}(j,j)}$ and $\sqrt{\mathbf{P}_l(j,j)}$ correspond to the standard deviations of the prior and posterior ensembles, respectively, for the $j$-th state component at iteration $l$.

Eq.~\eqref{eq:rtps} reinstates the reduced analysis spread toward the forecast spread and is constrained by $\beta_l^{(j)} \!\le\!\beta_{\max}$ to prevent over-inflation. By incorporating $\kappa = 0.7$ and $\beta_{\max}=1.2$~\cite{spratt2021characterizing}, we find that RTPS inflation effectively alleviates filter collapse and achieves more rapid and robust convergence compared to constant scalar inflation, as supported by numerical experiments.

While the RTPS method effectively maintains ensemble spread in nominal conditions, extreme initial biases can still lead to filter divergence, where the ensemble variance collapses before the mean migrates to the true solution. To further mitigate this failure mode, we introduce a strategy that selectively re-inflates the parameter subspace toward its initial relative ensemble spread when the filter exhibits concurrent signs of inconsistency and ensemble collapse. We refer to this strategy as ``parameter rejuvenation". Specifically, at iteration $l$ we monitor the normalized misfit
\begin{equation}
    \mathcal{J}_l
    = \frac{1}{N_y}\big(\bm{y}_{\text{obs}}-\mathbf{H}\bar{\bm{x}}_l\big)^\text{T}\mathbf{R}^{-1}\big(\bm{y}_{\text{obs}}-\mathbf{H}\bar{\bm{x}}_l\big),
    \label{eq:misfit}
\end{equation}
and the coefficient-of-variation (CV) vector of the unknown parameters, whose $j$-th component is defined as
\begin{equation}
    \boldsymbol{c}_l{(j)} = \frac{\sqrt{\mathbf{P}_l(j+N_y,j+N_y)}}{|\overline{\bm{x}}_l(j+N_y)|},
    \label{eq:CV}
\end{equation}
where $\sqrt{\mathbf{P}_l(j+N_y,j+N_y)}$ and $\overline{\bm{x}}_l(j+N_y)$ correspond to the standard deviation and mean of the $j$-th parameter. Rejuvenation is activated only when all three conditions below are satisfied simultaneously:

(1) \textbf{significant misfit}: the predicted output deviates substantially from the observations
\begin{equation}
    \mathcal{J}_l > \mathcal{J}_{\text{t}}, \quad \mathcal{J}_{\text{t}}=1.5;
    \label{eq:cond1}
\end{equation}

(2) \textbf{stagnation}: the relative decrease of misfit over a short window is small
\begin{equation}
    \frac{\mathcal{J}_{l-w}-\mathcal{J}_{l}}{\mathcal{J}_{l-w}} < \epsilon_{\text{d}},
    \quad w=3,\ \epsilon_{\text{d}}=0.1;
    \label{eq:cond2}
\end{equation}

(3) \textbf{ensemble spread collapse}: at least one calibrated parameter has collapsed CV
\begin{equation}
    \exists j:\quad \boldsymbol{c}_l{(j)}  < \eta_\text{c}\, \boldsymbol{c}_0{(j)},
    \quad \eta_\text{c}=0.5.
    \label{eq:cond3}
\end{equation}

When parameter rejuvenation is triggered, we bypass RTPS and instead apply a multiplicative re-inflation to the parameter ensemble
\begin{equation}
\bm{u}_l^{(i)} \leftarrow \bar{\bm{u}}_l + \alpha_{\mathrm r}\,\big(\boldsymbol{c}_0 \oslash \boldsymbol{c}_l\big)\odot\big(\bm{u}_l^{(i)}-\bar{\bm{u}}_l\big),
\qquad \alpha_{\mathrm r}=0.95,\qquad i=1,\ldots,N_{\mathrm e},
\label{eq:rejuv}
\end{equation}
where $\bar{\bm{u}}_l$ is the ensemble mean of the parameter vector, and $\oslash$ denotes Hadamard division.

\subsection{EnKF Inversion Framework}
\label{subsec:EnKFIF}

We have developed a non-intrusive, high-performance computational framework to implement the EnKF equations described in the preceding subsections. The framework is written in Python and executes the ensemble of forward solver calls required in the forecast step of each iteration in parallel using a fixed-size process pool. Specifically, we employ the \texttt{ProcessPoolExecutor} from Python’s \texttt{concurrent.futures} module, which provides a high-level interface for managing parallel function execution across multiple processors. The overall procedure is outlined in Algorithm~\ref{alg:enkf_inversion}, which presents a pseudocode implementation of this approach. Notably, at each EnKF iteration complete forward simulations are performed to obtain a time series of measurements (see Line~$9$ in Algorithm~\ref{alg:enkf_inversion}). Since the EnKF is both derivative-free and non-intrusive, the framework can be readily integrated with any forward solver treated as a black box.

\begin{algorithm}[H]
\setcounter{AlgoLine}{0}
\caption{\textbf{EnKF Inversion Framework}}
\label{alg:enkf_inversion}
\KwIn{initial guess of unknown parameters $\bm{u}_{0}$;
      initial error covariance $\mathbf{C}_0$;
      observation $\bm{y}_{\text{obs}}$;
      observation covariance $\mathbf{R}$;
      forward solver $\mathcal{G}$;
      ensemble size $N_{\text{e}}$;
      number of iterations $M$}

\textbf{Initialization}:

\For{$i=1,\dots,N_{\text{e}}$}{
    Generate the initial ensemble $\bm{u}_{0}^{(i)}$ using Eq.~\eqref{eq:initial_ensemble}\;
}

\For{$l=0,\dots,M$}{

  \textbf{Forecast}:
  
  \ForEachPar{$i=1,\dots,N_{\text{e}}$}{
     Compute the model output (i.e., the back-face deflection of the target plate) using the forward solver $\mathcal{G}\big(\bm{u}_l^{(i)}\big)$\;
  }
   
   Form the augmented state $\bm{x}_l$ (i.e., $\hat{\bm{x}}_{l+1}$) using Eq.~\eqref{eq:aug_state}\; 

   \textbf{Analysis}:
   
   Compute ensemble statistics $\overline{\bm{x}}_l$ and $\mathbf{P}_l$ using Eqs.~\eqref{eq:xbar} and \eqref{eq:Pforecast}\;

   Compute Kalman gain $\mathbf{K}_l$ using Eq.~\eqref{eq:enkf_gain}\; 
   
   Update the state $\bm{x}_{l+1}^{(i)},i=1,\dots,N_{\text{e}}$ using Eq.~\eqref{eq:enkf_update}\;
    
   \textbf{Covariance inflation}:

   Compute ensemble statistics $\overline{\bm{x}}_{l+1}$ and $\mathbf{P}_{l+1}$ using Eqs.~\eqref{eq:xbar} and \eqref{eq:Pforecast}\;

    Compute normalized misfit $\mathcal{J}_l$ using Eq.~\eqref{eq:misfit};

    Compute coefficient-of-variation $\boldsymbol{c}_l$ using Eq.~\eqref{eq:CV};

    \eIf{Eqs.~\eqref{eq:cond1}-\eqref{eq:cond3} are satisfied}{
        Update the parameters $\bm{u}_{l+1}^{(i)},i=1,\dots,N_{\text{e}}$ using Eq.~\eqref{eq:rejuv};
        }
        { 
        Compute multiplicative factor $\beta_{l+1}^{(j)}, j=1,\dots,N_y+N_u$ using Eq.~\eqref{eq:rtps}\;       
        Update the state $\bm{x}_{l+1}^{(i)},i=1,\dots,N_{\text{e}}$ using Eq.~\eqref{eq:infl_generic}\;
        }
    
   Extract updated parameters $\bm{u}_{l+1}^{(i)}$ from the augmented state $\bm{x}_{l+1}^{(i)},i=1,\dots,N_{\text{e}}$\;}
   
\KwOut{posterior ensemble $\bm{u}_{l}$ and
       posterior mean $\overline{\bm{u}}_{l}$ at all iterations $l=1,\dots,M$}
\end{algorithm}
\section{Comparison with a Traditional Bayesian Method}
\label{sec:comp}

We first compare the EnKF-based approach with a traditional sampling-based MCMC method. In general, MCMC requires on the order of $10^{4}$ (or more) sequential samples to obtain a well-resolved posterior, making it computationally prohibitive for HVI problems in which each forward SPH simulation is expensive. To provide a controlled comparison, we consider an impact-inspired ring-down problem with an analytical forward mapping. The ring-down problem describes the post-impact free vibration of a structural component (e.g., a plate) after the short-duration contact phase of a high-velocity impact has ended. We assume the ensuing response is dominated by the first bending mode and model the back-face deflection using a separable spatio-temporal representation. Because our goal is to calibrate model parameters using back-face deflection data, we define the forward operator $\mathcal{G}(\bm u)$ mapping the unknown parameters to the back-face deflection as
\begin{equation}
\mathcal G(\bm u)=\phi(x)\,q(\bm u;t),
\label{eq:toy_w}
\end{equation}
where $x\in[0,L]$ denotes a coordinate along a representative centerline. The spatial variation is represented by a simple symmetric mode shape
\begin{equation}
\phi(x)=\sin\!\left(\frac{\pi x}{L}\right).
\end{equation}
The temporal response follows an underdamped single-degree-of-freedom (SDOF) model
\begin{equation}
q(\bm u;t)=a\exp(-\zeta\omega_\text{n}t)\,\sin(\omega_\text{n}\sqrt{1-\zeta^2}t),
\label{eq:toy_q}
\end{equation}
parameterized by the unknown vector to be calibrated
\begin{equation}
\bm u = [a,\ \omega_\text{n},\ \zeta]^\text{T},
\end{equation}
where $a$ is the amplitude associated with impact severity, $\omega_\text{n}$ is the natural frequency governed by the effective stiffness-to-mass ratio, and $\zeta$ is the damping ratio capturing energy dissipation. This model yields an exponentially decaying sinusoidal deflection field that captures the essential ring-down behavior while avoiding explicit modeling of the contact phase and nonlinear impact dynamics.

The observation data consist of displacement measurements collected at $20$ equally spaced locations along the centerline (with $L=1$) and at three time instants $\{6\times10^{-4},\,1.0\times10^{-3},\,1.4\times10^{-3}\}\,$s, yielding a stacked observation dimension of $N_y=60$. The observations are generated using the true parameters
\(
\bm u_{\rm true}=\big[3.3~\text{mm},\ 2200~\text{rad}/\text{s},\ 0.06\big]^\text{T},
\)
and corrupted with additive Gaussian noise of zero mean and standard deviation $\sigma_{\eta}=1.0\times10^{-2}$~mm, i.e., $\bm y_{\rm obs}=\bm y_{\rm ref}+\bm\eta$ with $\bm\eta\sim\mathcal N(\bm 0,\sigma_\eta^2\bm I)$. For the EnKF simulations, we use an ensemble size of $N_e=100$ and run $M=100$ iterations. We also solve this example using a Random-Walk Metropolis MCMC (RWM-MCMC) scheme by running $3\times10^{4}$ steps and discarding the first $5\times10^{3}$ samples as burn-in; posterior samples are recorded every $10$ steps (thinning factor $10$). Both EnKF and RWM-MCMC are initialized using the same Gaussian prior in transformed coordinates $\bm\theta=[\log(a),\,\log(\omega_n),\,\mathrm{logit}(\zeta)]^\mathrm{T}$, where $\mathrm{logit}(\zeta)=\log\!\big(\zeta/(1-\zeta)\big)$. Specifically,
\(
\bm\theta\sim\mathcal N(\bm\theta_0,\mathrm{diag}(\bm\sigma_\theta^2))
\)
with $\bm\theta_0=[\log(2.0),\,\log(1800),\,\log(0.05/0.95)]^\mathrm{T}$ and $\bm\sigma_\theta=[0.35,\,0.25,\,0.60]^\mathrm{T}$.

\begin{figure}[htbp]
  \centering
  \includegraphics[width=0.95\textwidth]{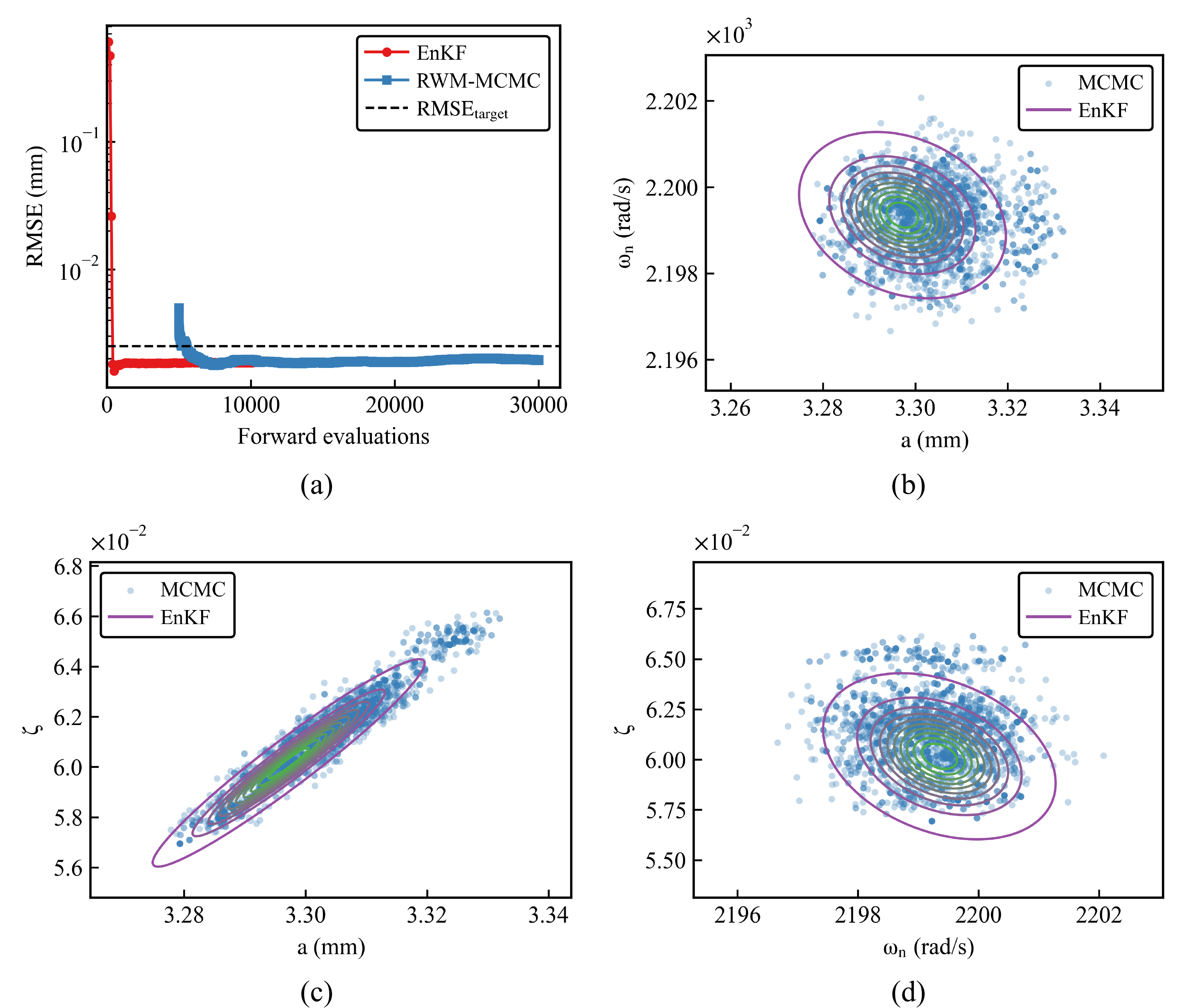} 
  \caption{Comparison of EnKF and RWM-MCMC for the ring-down model: (a) RMSE of the back-face displacement versus the number of forward evaluations. The dashed line indicates an illustrative target threshold $\mathrm{RMSE}$. (b)-(d) Pairwise posterior projections of the inferred parameter distributions. The EnKF results are shown as Gaussian-fit probability-mass contours at levels $\{0.10,0.20,0.30,0.40,0.50,0.60,0.70,0.80,0.90,0.99\}$.}
  \label{fig:enkf_vs_mcmc}
\end{figure}

Fig.~\ref{fig:enkf_vs_mcmc}(a) compares the evolution of the root-mean-square error (RMSE) of the back-face displacement at the selected locations and time instants as a function of the number of forward evaluations. A more detailed comparison of computational cost is provided in Table~\ref{tab:enkf_vs_mcmc_results}. Because burn-in contributes to the wall-clock cost of sequential MCMC, it is included when comparing the computational cost of RWM-MCMC with EnKF. As shown, the EnKF converges within the first $4$ iterations, requiring $500$ forward evaluations, whereas RWM-MCMC requires approximately $6000$ iterations (and forward evaluations) to reach a comparable level of accuracy. The final RMSE obtained by EnKF is $1.86\times10^{-3}$~mm, which is comparable to that of RWM-MCMC ($1.95\times10^{-3}$~mm), indicating that both methods recover posterior means that are close to the true values. To quantify computational efficiency in a fair and interpretable manner, we introduce a Cost-to-Misfit metric, $\mathcal{C}$, defined as the number of forward evaluations required to reduce the RMSE below a prescribed threshold. For example, for a target RMSE of $2.5\times10^{-3}$~mm (Fig.~\ref{fig:enkf_vs_mcmc}(a)), RWM-MCMC requires $5512$ forward evaluations, while EnKF requires only $400$, i.e., an approximately $13.8\times$ reduction in forward evaluations (at least one order of magnitude).

\begin{table}[htbp]
\centering
\caption{Efficiency comparison on the ring-down model.}
\label{tab:enkf_vs_mcmc_results}
\setlength{\tabcolsep}{8pt}
\renewcommand{\arraystretch}{1.2}
\begin{tabular}{lcccc}
\toprule
Method & Type & Total Evals. & $\mathcal{C}(\mathrm{RMSE}_{\rm target})$ & Final RMSE (mm) \\
\midrule
EnKF & Parallel & 10,100 & 400 & $1.86\times 10^{-3}$ \\
RWM-MCMC & Sequential & 30,000 & 5,512 & $1.95\times 10^{-3}$ \\
\bottomrule
\end{tabular}
\end{table}

Figs.~\ref{fig:enkf_vs_mcmc}(b)-(d) compare the pairwise posterior distributions obtained from EnKF with those from RWM-MCMC. The two sets of posteriors are in close agreement, indicating that EnKF recovers a posterior distribution comparable to that of RWM-MCMC, but at a substantially lower computational cost.

\section{Application to HVI}
\label{sec:nexperiments}

To demonstrate the application of the EnKF framework, we consider the characterization of material models in a representative HVI scenario. The target is a rectangular AZ31B magnesium alloy plate with dimensions of $35.56 \,\text{mm} \times 35.56 \,\text{mm} \times 9.5 \,\text{mm}$, selected for its relevance as a lightweight structural material in aerospace applications. The projectile is a steel sphere with a diameter of $5 \,\text{mm}$, impacting the plate at a velocity of $1.2 \,\text{km/s}$ under normal incidence. This configuration provides a well-controlled setup for evaluating the predictive capability of the material models and the effectiveness of the proposed data assimilation approach.

\subsection{Forward Solver}
\label{subsec:FS}
\begin{figure}[!htbp]
\centering
\includegraphics[width=0.6\textwidth]{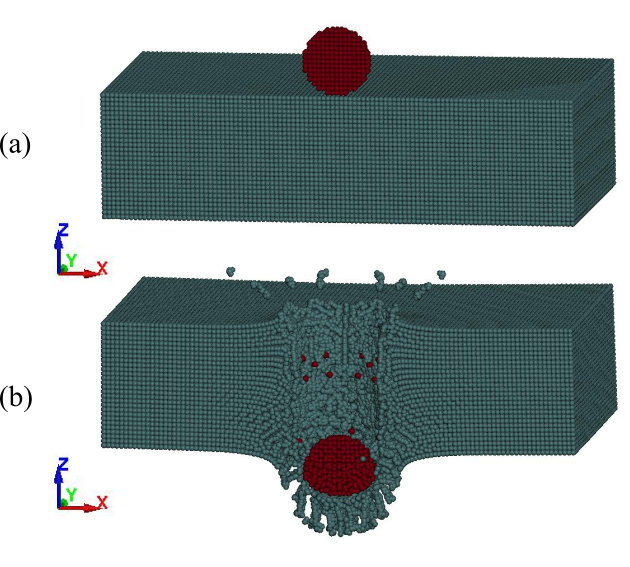}
\caption{Schematic illustration of the SPH model of the HVI problem with half the geometry hidden: (a) Initial state $t=0~\upmu\rm{s}$, and (b) Termination state $t=20.0~\upmu\rm{s}$. Red and green markers represent SPH particles of the projectile and the plate, respectively}
\label{fig:SPH_model}
\end{figure}

The numerical simulation of the HVI problem is performed using the SPH solver implemented in the commercial software LS-DYNA (version R14.1)~\cite{lsdynatheory}. Both the projectile and the target plate are discretized with SPH particles. To achieve a balance between accuracy and computational cost, the target plate is discretized with 320,572 particles, while the projectile is discretized with 1,791 particles. To further improve numerical precision, particles are arranged in a uniformly distributed configuration to minimize discrepancies in inter-particle spacing~\cite{lsdynatheory}. The initial SPH geometry of the plate and projectile is shown in Fig.~\ref{fig:SPH_model}(a). Particle interactions are modeled using the cubic spline kernel approximation in Eq.~(\ref{eq:cubicspline}), while contact between the projectile and target is automatically handled by the neighboring particle search, eliminating the need for user-defined contact algorithms~\cite{lsdynatheory}.  

All simulations are terminated at $20.0~\upmu\text{s}$, at which time the maximum back-face deflection of the plate exceeds the projectile diameter. This termination point ensures the availability of deflection data of sufficient magnitude for material model characterization. The final configuration of the system is illustrated in Fig.~\ref{fig:SPH_model}(b). An adaptive time step, determined by the critical particle size, is employed throughout the simulations. Heat transfer effects are neglected under the assumption that thermal conduction occurs on a much slower timescale than the impact process, such that the calculations are performed adiabatically with an initial temperature set to room temperature.

The projectile material is Firth-Vickers 535 (FV535) martensitic stainless steel~\cite{erice2012flow}, and the target plate material is the AZ31B magnesium alloy~\cite{hasenpouth2010tensile}. The constitutive and failure responses of both materials are modeled using the JC formulation given in Eqs.~\eqref{eq:JCplasticity}-\eqref{eq:JCfractureepsilon}. For the target plate, the JC plasticity and fracture parameters are treated as uncertain and require calibration. For reference, the baseline values of these parameters are provided in Table~\ref{tab:AZ31BJCEOSpara}, while the fixed material parameters of the plate are summarized in Table~\ref{tab:AZ31Bnominalpara}. The projectile material model is assumed to be well characterized, and its baseline properties are listed in Table~\ref{tab:fv535matpara}. 
The volumetric response of the projectile is characterized using a linear polynomial EOS, where the bulk elastic modulus serves as the sole coefficient.

\begin{table}[!ht]
\centering
\caption{Johnson-Cook and Grüneisen EOS parameters for AZ31B magnesium alloy target.}
\begin{tabularx}{0.8\linewidth}{X L{2.5cm} L{1.5cm} L{1.5cm}}
\toprule
\toprule
Parameter & Value & Unit & Source \\
\midrule
Johnson-Cook plasticity $A$ & $225.171$ & [MPa] & \cite{hasenpouth2010tensile} \\
Johnson-Cook plasticity $B$ & $168.346$ & [MPa] & \cite{hasenpouth2010tensile} \\
Johnson-Cook plasticity $n$ & $0.242$   & -   & \cite{hasenpouth2010tensile} \\
Johnson-Cook plasticity $C$ & $0.013$   & -   & \cite{hasenpouth2010tensile} \\
Johnson-Cook plasticity $m$ & $1.55$   & -   & \cite{hasenpouth2010tensile} \\
Johnson-Cook fracture $D_1$      & $-0.35$   & -   & \cite{feng2014constitutive} \\
Johnson-Cook fracture $D_2$      & $0.6025$  & -   & \cite{feng2014constitutive} \\
Johnson-Cook fracture $D_3$      & $-0.4537$   & -   & \cite{feng2014constitutive}\\
Johnson-Cook fracture $D_4$      & $0.4738$   & -   & \cite{sun2022uncertainty} \\
Johnson-Cook fracture $D_5$      & $7.2$   & -   & \cite{feng2014constitutive} \\
Grüneisen EOS $C_s$    & $4520.0$  & [m/s]   & \cite{green1989reaction}\\
Grüneisen EOS $S_1$         & $1.242$   & -   & \cite{green1989reaction}\\
Grüneisen EOS $\gamma_0$    & $1.54$    & -   & \cite{green1989reaction}\\
Grüneisen EOS $\alpha$      & $0.33$    & -   & \cite{green1989reaction}\\
\bottomrule
\end{tabularx}
\label{tab:AZ31BJCEOSpara}
\end{table}

\begin{table}[!ht]
\centering
\caption{Nominal mechanical properties for AZ31B magnesium alloy target.}
\begin{tabularx}{0.8\linewidth}{X L{2.5cm} L{1.5cm} L{1.5cm}}
\toprule
\toprule
Parameter & Value & Unit & Source \\
\midrule
Reference mass density & $1.77$ & [g/cm$^3$] & \cite{hasenpouth2010tensile} \\
Young's Modulus & $45.0$ & [GPa] & - \\
Poisson's ratio & $0.35$   & -   & - \\
Specific heat & $1.005$   & [J/(K·g)]  & \cite{lee2013thermal} \\
Taylor-Quinney factor & $0.6$   & -   & \cite{kingstedt2019conversion} \\
Reference strain rate     & $0.001$   & [s$^{-1}$]   & \cite{hasenpouth2010tensile} \\
Reference Temperature      & $298.15$  & [K]  & \cite{hasenpouth2010tensile} \\
Reference melting Temperature      & $905$   & [K]   & \cite{hasenpouth2010tensile}\\
\bottomrule
\end{tabularx}
\label{tab:AZ31Bnominalpara}
\end{table}

\begin{table}[!htbp]
\centering
\caption{Material parameters for FV535 stainless steel projectile.}
\begin{tabularx}{0.8\linewidth}{X L{2.5cm} L{1.5cm} L{1.5cm}}
\toprule
\toprule
Parameter & Value & Unit & Source \\
\midrule
Reference Mass density & $7.85$ & [g/cm$^3$] & \cite{erice2012flow} \\
Young's modulus & $210$ & [GPa] & \cite{erice2012flow} \\
Poisson's ratio & $0.28$ & - & \cite{erice2012flow} \\
Specific heat & $0.46$ & [J/(K·g)] & \cite{erice2012flow} \\
Reference melting temperature & $1053$ & [K] & \cite{erice2012flow} \\
Johnson-Cook plasticity $A$ & $1035$ & [MPa] & \cite{erice2012flow} \\
Johnson-Cook plasticity $B$ & $190$ & [MPa] & \cite{erice2012flow} \\
Johnson-Cook plasticity $n$ & $0.3$   & -   & \cite{erice2012flow} \\
Johnson-Cook plasticity $C$ & $0.006$   & -   & \cite{erice2012flow} \\
Johnson-Cook plasticity $m$ & $4.5$   & -   & \cite{erice2012flow} \\
Johnson-Cook fracture $D_1$      & $0.1133$   & -   & \cite{erice2012flow} \\
Johnson-Cook fracture $D_2$      & $2.11$  & -   & \cite{erice2012flow} \\
Johnson-Cook fracture $D_3$      & $-1.65$   & -   & \cite{erice2012flow}\\
Johnson-Cook fracture $D_4$      & $0.0125$   & -   & \cite{erice2012flow} \\
Johnson-Cook fracture $D_5$      & $0.9768$   & -   & \cite{erice2012flow} \\
\bottomrule
\end{tabularx}
\label{tab:fv535matpara}
\end{table}

\subsection{HVI Behavior}
\label{subsec:HVIB}
\begin{figure}[!htbp]
\centering
\includegraphics[width=0.8\textwidth]{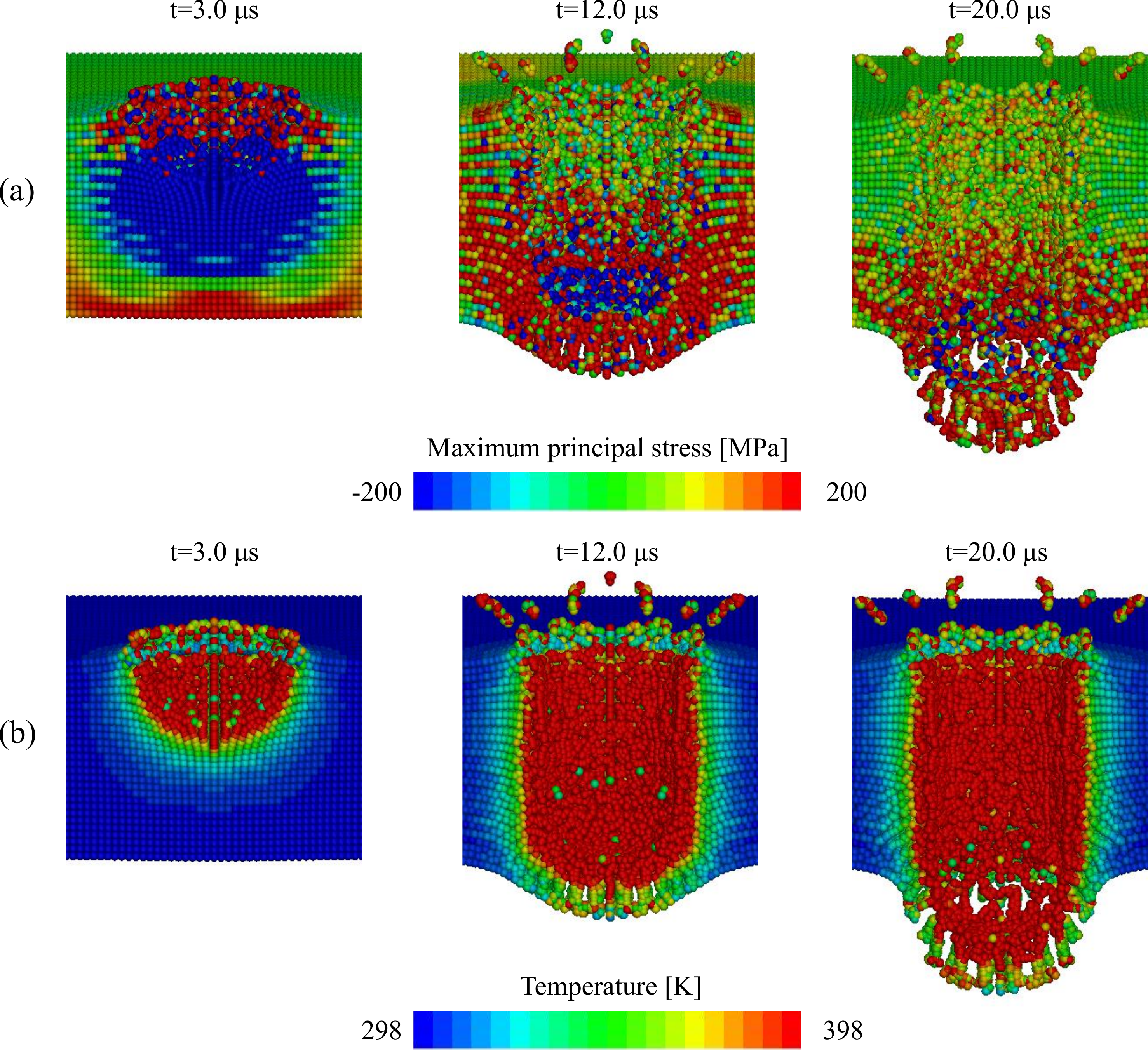}
\caption{Numerical simulation results. (a) Maximum principal stress contours of the target plate. (b) Temperature field contours of the target plate.}
\label{fig:mps_temp}
\end{figure}

In this subsection, we evaluate the HVI response of the plate using the baseline material parameters listed in Tables~\ref{tab:AZ31BJCEOSpara}-\ref{tab:fv535matpara}. The objective is to assess the fidelity and predictive capability of the SPH method. Fig.~\ref{fig:mps_temp} presents the level contours of the maximum principal stress and temperature at three representative time instants: $t = 3.0~\upmu\text{s}$, $12.0~\upmu\text{s}$, and $20.0~\upmu\text{s}$. For clarity of presentation, the projectile and half of the target plate are removed during post-processing to enable an unobstructed view of field evolution within the contact region.  

At $t = 3.0~\upmu\text{s}$, Fig.~\ref{fig:mps_temp}(a) shows the generation of a compressive shock wave at the impact interface, which propagates through the thickness of the target plate~\cite{kawai2015stress}. This compressive wave elevates local pressure and stress. Upon reaching the free back-face of the plate, the shock reflects as a tensile rarefaction wave, which propagates inward and interacts with a secondary tensile wave reflected from the projectile. Their superposition produces a high-magnitude tensile core that can initiate spallation~\cite{meyers1983dynamic}. At the same time, Fig.~\ref{fig:mps_temp}(b) indicates a local temperature increase exceeding $100$ K near the developing cavity, due to the conversion of plastic work into heat.  

By $t = 12.0~\upmu\text{s}$, Figs.~\ref{fig:mps_temp}(a) and (b) show that perforation occurs primarily through a conventional plugging mechanism~\cite{klepaczko1998remarks}. The impact energy of the projectile is mainly dissipated as plastic work concentrated around shear rupture zones that isolate the projectile from the plate. The temperature field also peaks at the cavity boundary, producing thermal softening that promotes deformation localization and accelerates plug formation.  

At $t = 20.0~\upmu\text{s}$, the shear plug has fully perforated the plate. Residual tensile stresses drive lateral cracks parallel to the midplane, producing a characteristic disk-type lamination~\cite{fras2020modeling}. The simulation further reveals fragmented material distributed on both the front- and back-faces of the plate, a typical feature of materials subjected to HVI.  

\subsection{Synthetic Observations}
\label{subsec:SOD}

\begin{figure}[!htbp]
\centering
\includegraphics[width=0.65\textwidth]{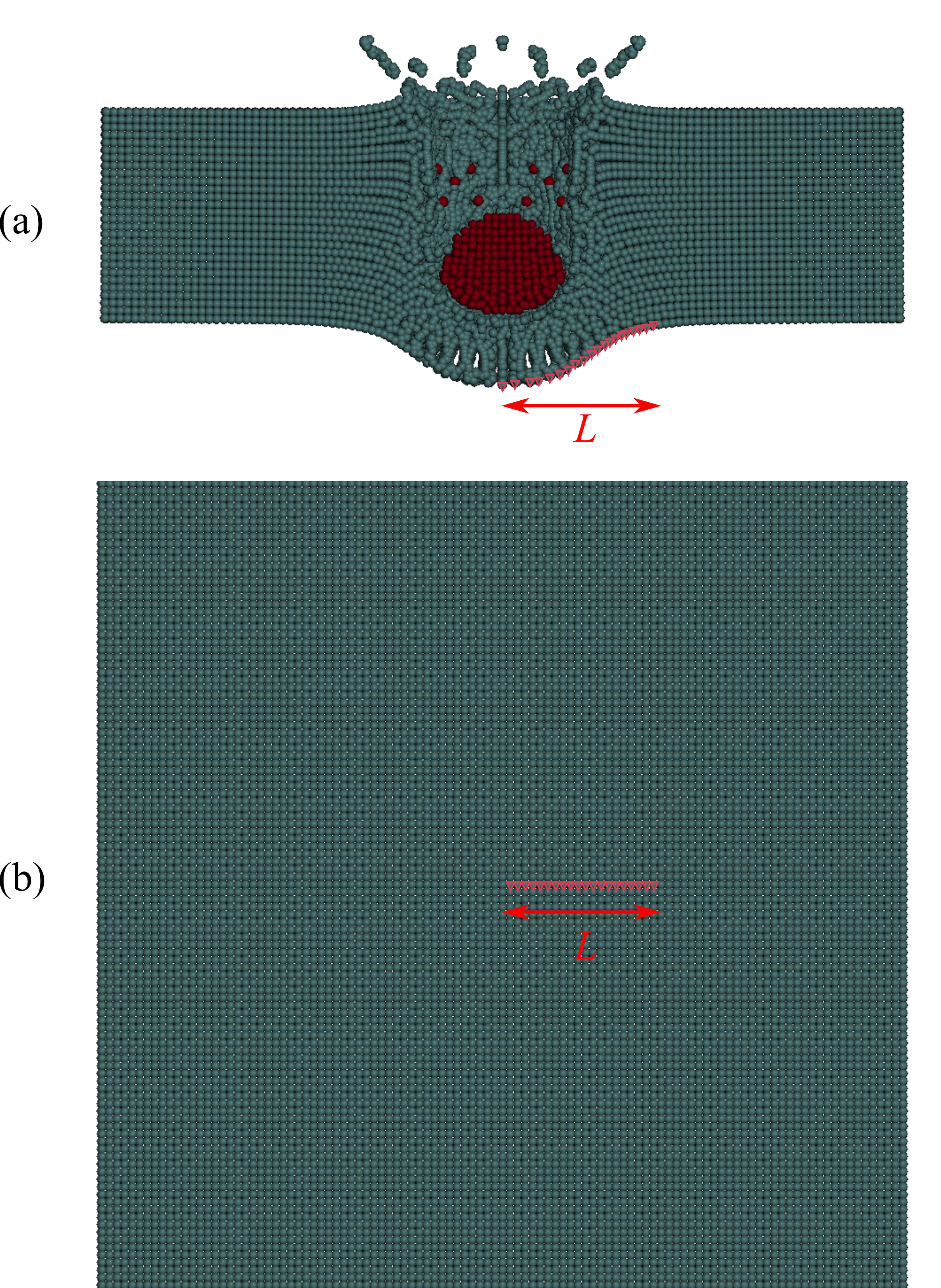}
\caption{Node set selected for observation in the numerical experiment, highlighted by inverted red triangles: (a) side view, and (b) back view. Red and green markers represent SPH particles of the projectile and the plate, respectively.}
\label{fig:nodset_obs}
\end{figure}

In the numerical experiment, we assume that the back-face deflection of the target plate can be measured during the HVI process. Due to the very high impact velocity and the short duration of the event, the periphery of the plate remains nearly stationary, and the particles near the periphery exhibit negligible displacement. Therefore, we focus on particles located in a high-motion zone surrounding the impact region and use their time-series displacement data to characterize the material models. The locations of these particles are shown in Fig.~\ref{fig:nodset_obs}. The observation line has a length of approximately $L = 6.6$~mm from the center of the impact zone, which exceeds the projectile diameter of $5$~mm, and includes $N_\text{o} = 20$ observed particles. In addition, we have performed a sensitivity study to examine how the observation-line length influences the characterization results. These findings will be discussed later.

We use synthetic observations to mimic the data that would be measured experimentally using DGS or 3D-DIC. However, the reliability of 3D-DIC is limited to the early stages of penetration, as gross fracture and the formation of a debris cloud obscure the speckle pattern and cause correlation failure~\cite{malhotra2025modified}. To avoid this experimental constraint, synthetic observations are extracted at three early time instants where the displacement field remains both measurable and dynamically significant, namely $t=10.0~\upmu\text{s}$, $11.0~\upmu\text{s}$, and $12.0~\upmu\text{s}$. The minimum and maximum true deflections at these times, evaluated over the selected SPH particles, are summarized in Table~\ref{tab:dimensions}. The corresponding maximum deflections are $1.98$~mm, $2.47$~mm, and $2.98$~mm, which are on the order of projectile diameter. Accordingly, the observation $\bm{y}$ consists of the displacement in the impact direction of the $N_\text{o}=20$ observed particles at the three time instants, giving an observation vector of dimension $N_y = 3N_\text{o} = 60$.

\begin{table}[!htbp]
\centering
\caption{Minimum and maximum true values of measured back-face deflection at the three selected time instants.}
\begin{tabularx}{0.8\linewidth}{L{8cm}XX}
\toprule
\toprule
Part & Value & Unit \\
\midrule
Projectile diameter & $5$ & [mm] \\
Target thickness & $9.5$ & [mm] \\
Minimum back-face deflection at $t=10.0~\upmu\rm{s}$ & $0.25$ & [mm] \\
Minimum back-face deflection at $t=11.0~\upmu\rm{s}$ & $0.29$ & [mm] \\
Minimum back-face deflection at $t=12.0~\upmu\rm{s}$ & $0.32$ & [mm] \\
Maximum back-face deflection at $t=10.0~\upmu\rm{s}$ & $1.98$ & [mm] \\
Maximum back-face deflection at $t=11.0~\upmu\rm{s}$ & $2.47$ & [mm] \\
Maximum back-face deflection at $t=12.0~\upmu\rm{s}$ & $2.98$ & [mm] \\
\bottomrule
\end{tabularx}
\label{tab:dimensions}
\end{table}

The synthetic observation $\bm{y}_{\rm obs}$ is generated according to
\begin{equation}
\label{eq:obs_}
    \bm{y}_{\rm obs} = \bm{y}_{\rm true} + \boldsymbol{\eta}_0,
\end{equation}  
where $\bm{y}_{\rm true}$ represents the model output at the baseline parameter values. The term $\boldsymbol{\eta}_0$ denotes a systematic bias, set to $-0.001$~mm. The observation covariance $\mathbf{R}$ is assumed to be diagonal, and all measured deflections share the same standard deviation. This standard deviation is assumed to be a fixed value, which is estimated by multiplying the reported random error $4\%$ of 3D-DIC by the smallest observable displacement listed in Table~\ref{tab:dimensions}, yielding a value of $0.01$~mm. This procedure produces a dataset that captures both the spatio-temporal distribution and noise characteristics of typical 3D-DIC measurements, thereby providing a realistic testing environment for evaluating the EnKF framework. 

\subsection{Input Space Reduction}
\label{subsec:MIO}

As described in Section~\ref{sec:method}, we calibrate the JC plasticity, JC fracture, and Grüneisen EOS models of the plate used in the SPH simulations. In total, this involves $14$ parameters: $A$, $B$, $n$, $C$, and $m$ in the JC plasticity model; $D_1$, $D_2$, $D_3$, $D_4$, and $D_5$ in the JC fracture model; and $C_s$, $S_1$, $\gamma_0$, and $\alpha$ in the Grüneisen EOS model. Calibrating all $14$ parameters simultaneously would require the EnKF to operate in a high-dimensional state space, where limited ensemble sizes lead to rank-deficient covariances and incorrect long-range correlations that can significantly degrade filter performance~\cite{ruiz2013estimating}.  

To mitigate this issue while validating the EnKF inversion framework, we reduce the dimensionality of the unknown parameter space through an initial screening. Specifically, we perform a one-at-a-time (OAT) sensitivity analysis~\cite{borgonovo2016sensitivity}, in which each parameter is perturbed individually while the others remain fixed at their baseline values. The output of this OAT analysis is the noise-free back-face deflection response $\bm{y}_j$ ($j=1,2,3$) at the three time instants $t = 10.0~\upmu\rm{s}$, $11.0~\upmu\rm{s}$, and $12.0~\upmu\rm{s}$, as described in Section~\ref{subsec:SOD}. Local sensitivity is quantified by the root-mean-square deviation (RMSD) between the positively and negatively perturbed simulations
\begin{equation}
    S_{i,j} = \sqrt{\frac{1}{N_{y_j}}
        \bigl\lvert
        \bm{y}_j(\hat{\bm{\theta}}_i, \theta_i^+) -
        \bm{y}_j(\hat{\bm{\theta}}_i, \theta_i^-)
        \bigr\rvert ^2},
    \quad i = 1,\dots,N_\theta,~j=1,2,3,
\end{equation}
where
\begin{subequations}
\begin{align}
 (\hat{\bm{\theta}}_i, \theta_i^+) &= (\theta_1,\ldots,\theta_{i-1}, \theta_i + \delta_i,\theta_{i+1},\ldots,\theta_{N_\theta}), \\
 (\hat{\bm{\theta}}_i, \theta_i^-) &= (\theta_1,\ldots,\theta_{i-1}, \theta_i - \delta_i,\theta_{i+1},\ldots,\theta_{N_\theta}),
\end{align}
\end{subequations}
and $\delta_i$ is the perturbation magnitude. In this analysis, the vector of parameters is
\begin{equation}
\bm{\theta}
  = [A, B, n, C, m, D_1, D_2, D_3, D_4, D_5, C_s, S_1, \gamma_0, \alpha]^{\mathrm T},
\end{equation}
with $N_\theta = 14$. Each parameter $\theta_i$ is perturbed by $\pm 25\%$ of its baseline value, i.e., $\delta_i = 0.25\,\theta_i$.

\begin{figure}[!htbp]
\centering
\includegraphics[width=0.9\textwidth]{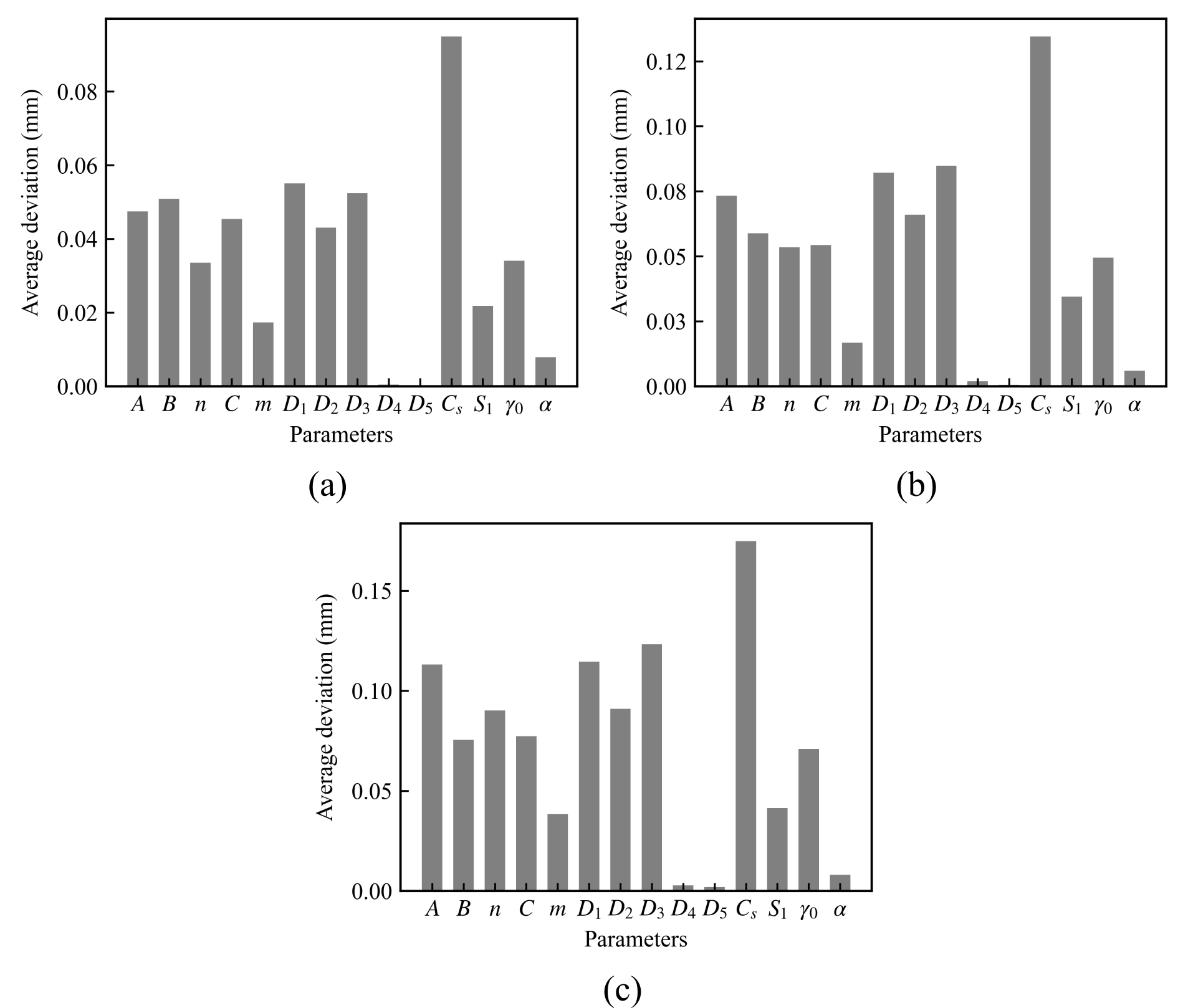}
\caption{RMSD of back-face deflection computed at (a) $t = 10.0\,\upmu\rm{s}$, (b) $t = 11.0\,\upmu\rm{s}$, and (c) $t = 12.0\,\upmu\rm{s}$.}
\label{fig:back-face_rmse}
\end{figure}

The results of the OAT sensitivity analysis are presented in Fig.~\ref{fig:back-face_rmse}. The JC fracture parameters $D_4$ and $D_5$ exhibit nearly negligible sensitivity at all three observation times, indicating minimal influence on the back-face deflection. For model calibration, we exclude $D_5$ but retain $D_4$ to demonstrate how the EnKF framework characterizes parameters with low sensitivity.  

We further remove the JC quasi-static parameters $A$, $B$, $n$, $D_1$, $D_2$, and $D_3$, which can be calibrated straightforwardly through uniaxial tension tests. The temperature-related JC plasticity parameter $m$ is also excluded to reduce input space. Among the Grüneisen EOS constants, $C_s$, $S_1$, and $\alpha$ are well tabulated for Mg alloys~\cite{marsh1980lasl}. However, the parameter $\gamma_0$ shows order-of-magnitude variability across different references~\cite{green1989reaction,yuan2017numerical,altair_help2025} and is therefore retained.  

Guided by the OAT analysis and the practical considerations of experimental calibration, we ultimately restrict the inversion to a reduced three-dimensional parameter set, selecting one parameter from each material model. Accordingly, the parameter vector for the EnKF calibration is defined as
\begin{equation}
\label{eq:model_in}
    \bm{u} = [C,\; D_4,\; \gamma_0 ]^{\mathrm T},
\end{equation}
with $N_u = 3$, while all remaining coefficients are fixed at their baseline values. This selection preserves the parameters that are both the most difficult to measure experimentally and the most influential on the dynamic response, thereby improving the identifiability and improving the stability of the filter.

\begin{figure}[!htbp]
\centering
\includegraphics[width=0.9\textwidth]{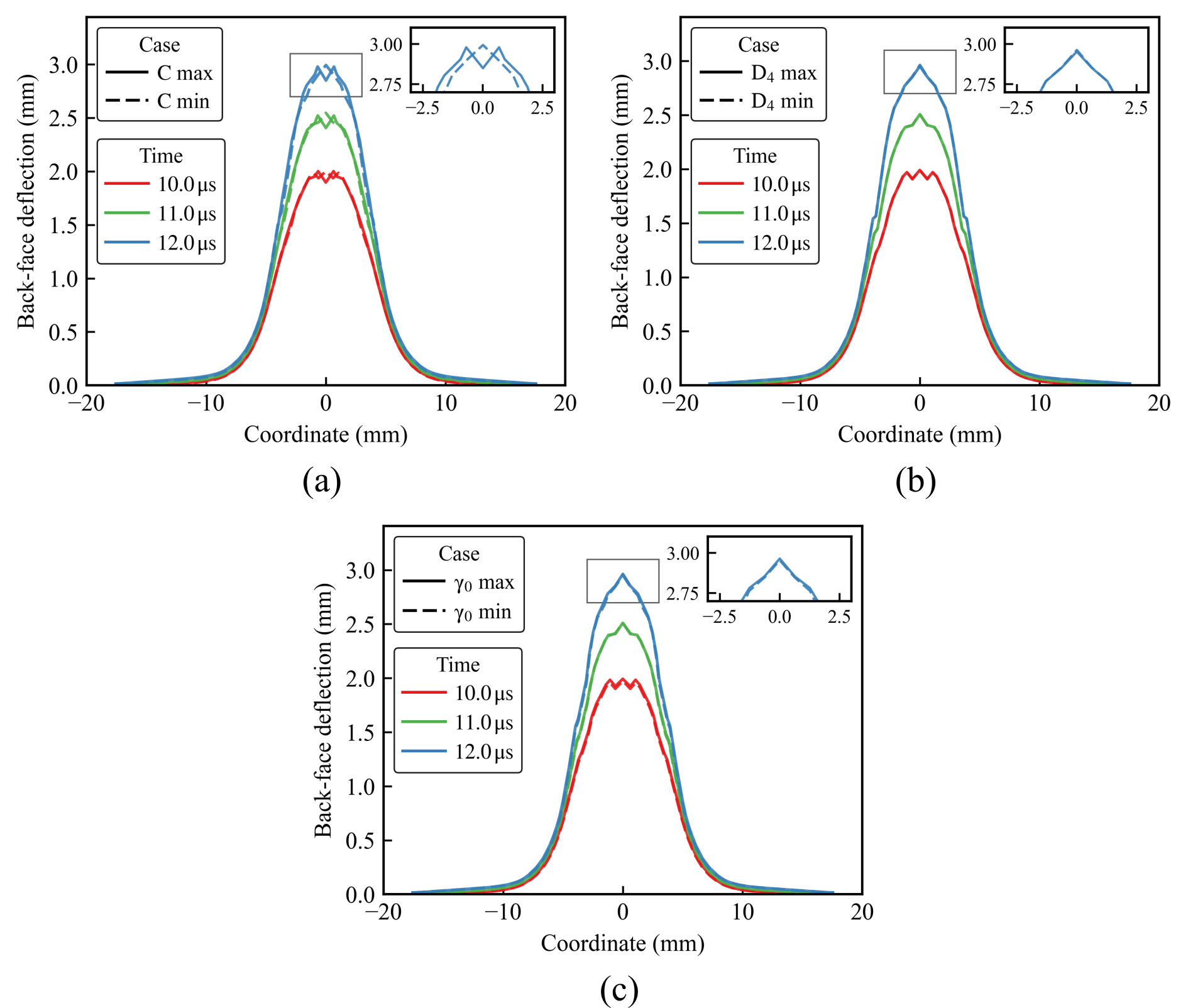}
\caption{Back-face deflection profiles produced by the extreme perturbations $\theta_i^{-}=0.75\,\theta_i$ (dashed) and $\theta_i^{+}=1.25\,\theta_i$ (solid) of the retained parameters: (a) $C$, (b) $D_4$, and (c) $\gamma_0$. Curves are colored by time, red: $t = 10.0\,\upmu\rm{s}$; green: $t = 11.0\,\upmu\rm{s}$; blue: $t = 12.0\,\upmu\rm{s}$.}
\label{fig:back-face_xzplots}
\end{figure}

To further illustrate the influence of $C$, $D_4$, and $\gamma_0$ on the observable response, Fig.~\ref{fig:back-face_xzplots} shows the back-face deflection for the lower perturbation $(\theta_i^{-}=0.75\,\theta_i;\ \text{dashed})$ and the upper perturbation $(\theta_i^{+}=1.25\,\theta_i;\ \text{solid})$ of each retained parameter. Among the three, varying $C$ produces the largest spread in the deflection curve, whereas $\gamma_0$ mainly induces a nearly uniform vertical shift, and the curve associated with $D_4$ is barely distinguishable from the baseline. Notably, the deflection curves of $C$ and $\gamma_0$ corresponding to the lower and upper perturbations still remain very close to each other, indicating that the back-face deflection is only weakly sensitive to these parameters. Nevertheless, as shown in the next section, our EnKF framework still recovers accurate estimates of the model parameters, underscoring the robustness of the Bayesian calibration procedure for practical applications.

\subsection{EnKF Inversion Results}
\label{subsec:Results}

We employ the developed EnKF inversion framework to calibrate the model parameters $C$, $D_4$, and $\gamma_0$. The initial guess for the unknown parameters is taken as the mean of the initial ensemble $\bm{u}_0$. The prior covariance matrix $\mathbf{C}_0$ is assumed diagonal, with the standard deviation of each component set to $10\%$ of its initial guess.

To evaluate the performance of the EnKF inversion framework described in Algorithm~\ref{alg:enkf_inversion}, we consider four scenarios that differ in the initial guess $\bm{u}_0$ and in the amount of synthetic observation data:
\begin{enumerate}
    \item Case 1: under-biased initial guess, $\bm{u}_0 = 0.75\,\bm{u}_{\text{true}}$, where $\bm{u}_{\text{true}}$ denotes the baseline (true) parameter values;
    \item Case 2: over-biased initial guess, $\bm{u}_0 = 1.25\,\bm{u}_{\text{true}}$;
    \item Case 3: same initial guess as Case~1 but with fewer observations, i.e., $N_{\rm o}=10$ and an observation line of length $L=3.3$~mm from the center of the impact zone.
    \item Case 4: strongly biased initial guess, $\bm{u}_0 = 2.5\,\bm{u}_{\text{true}}$, where the initial guesses for all parameters deviate by $150\%$ from their true values;

\end{enumerate}

\begin{figure}[htbp]
  \centering
  \includegraphics[width=0.95\textwidth]{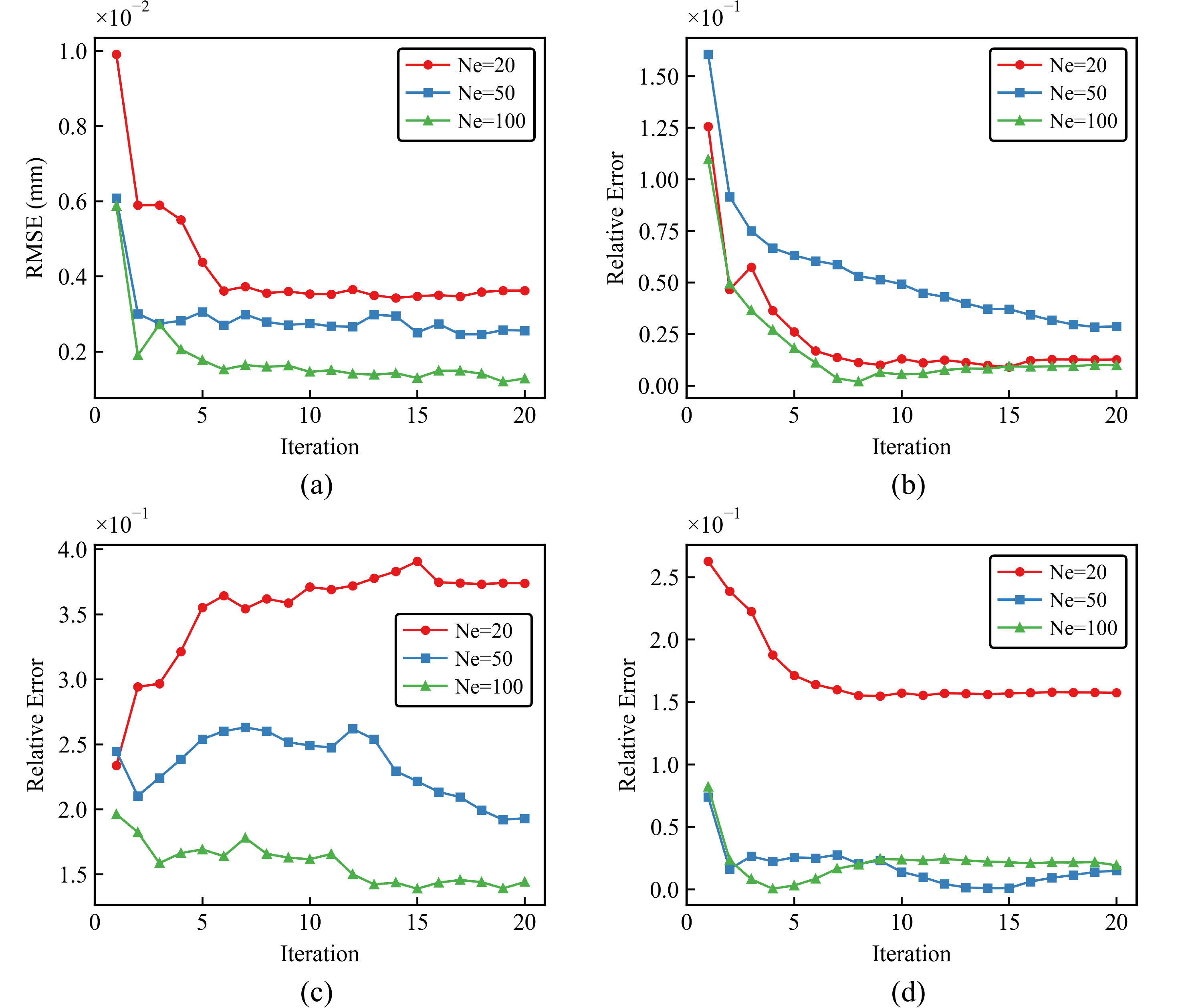}
  \caption{Convergence histories for ensemble sizes $N_{\mathrm e}\in\{20,50,100\}$ using the Case~1 setup: (a) RMSE of the back-face deflection. (b)-(d) Relative errors of the posterior-mean estimates for the calibrated parameters $C$, $D_4$, and $\gamma_0$, respectively.}
  \label{fig:convergence_Ne}
\end{figure}

Before presenting the detailed calibration results, we first justify the choice of ensemble size $N_\text{e}$ and the number of EnKF iterations $M$ using the representative setup of Case~1. To assess convergence, we compute at each EnKF iteration the RMSE of the back-face deflection relative to the ground-truth response and the relative error of the calibrated parameters with respect to their true values. Fig.~\ref{fig:convergence_Ne} summarizes the results for three ensemble sizes, $N_\text{e}\in\{20,50,100\}$, with all other settings identical to Case~1. As shown, smaller ensembles lead to noticeably larger errors in both the back-face deflection and the inferred parameters. In particular, the RMSE of the back-face deflection is approximately $4.17\times10^{-3}$~mm for $N_\text{e}=20$, whereas it decreases to $7.77\times10^{-4}$~mm for $N_\text{e}=100$. This improvement is expected, as larger ensembles provide a more accurate representation of the forecast covariance and more adequate exploration of the parameter space. Accordingly, we use $N_{\mathrm e}=100$ in the following studies. With this choice, the EnKF converges rapidly, reaching a stable solution within roughly the first $5$-$8$ iterations. Although most of the data-fit improvement occurs early, we adopt $M=20$ as a conservative upper bound to ensure robustness in cases with large initial biases (e.g., Case~4).

\begin{figure}[!htbp]
\centering
\includegraphics[width=0.95\textwidth]{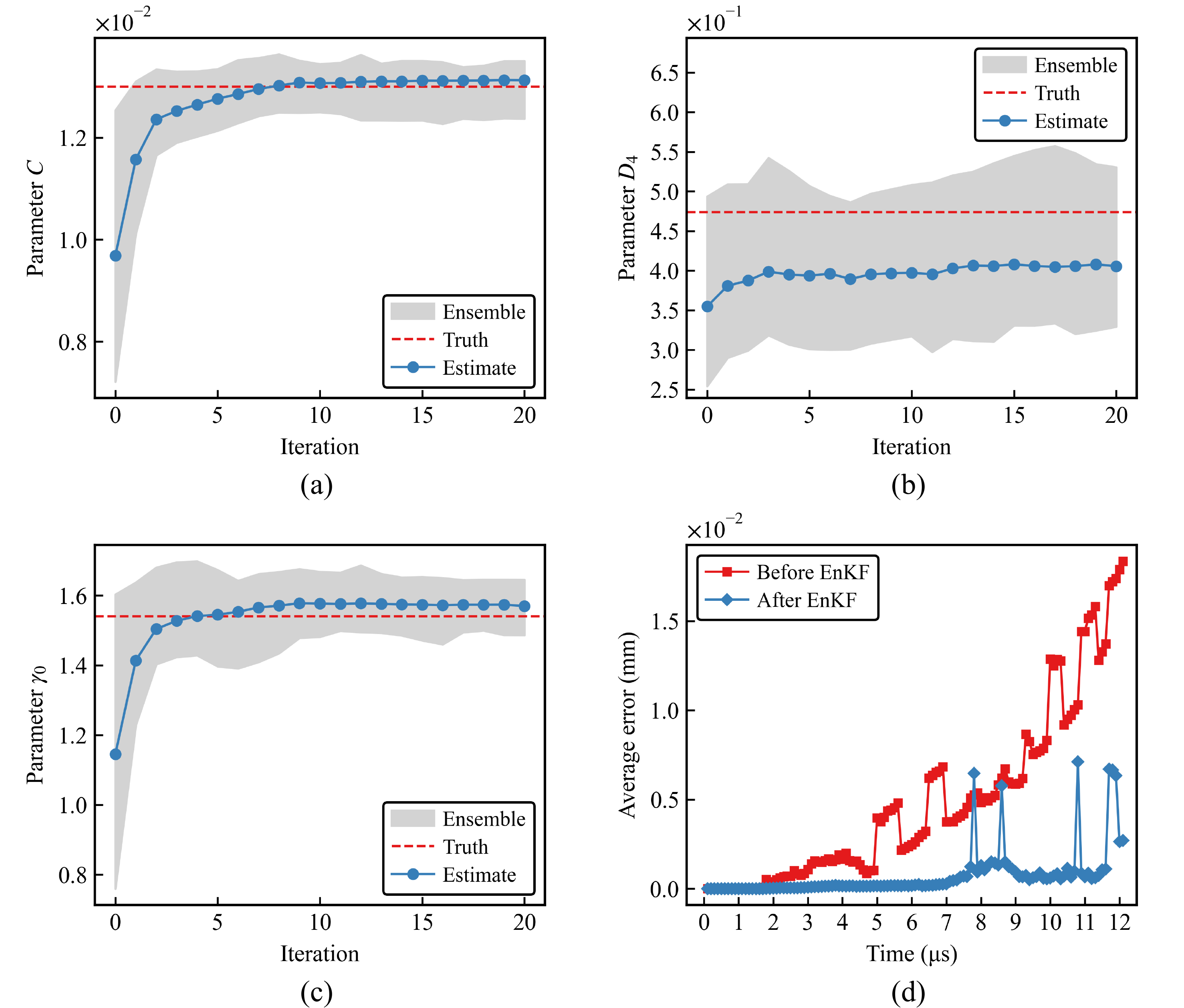}
\caption{EnKF results for Case 1 with initial ensemble mean $\bm{u}_0 = 0.75 \, \bm{u}_{\text{true}}$ ($-25\%$ bias): (a)-(c) Iteration histories of inferred parameters $C$, $D_4$, and $\gamma_0$. (d) Time-resolved RMSE of the back-face deflection before and after assimilation with respect to the true deflection over the $0$-$12 \, \upmu\rm{s}$ observation window.}
\label{fig:EnKF_result_1}
\end{figure}

\begin{figure}[!htbp]
\centering
\includegraphics[width=0.95\textwidth]{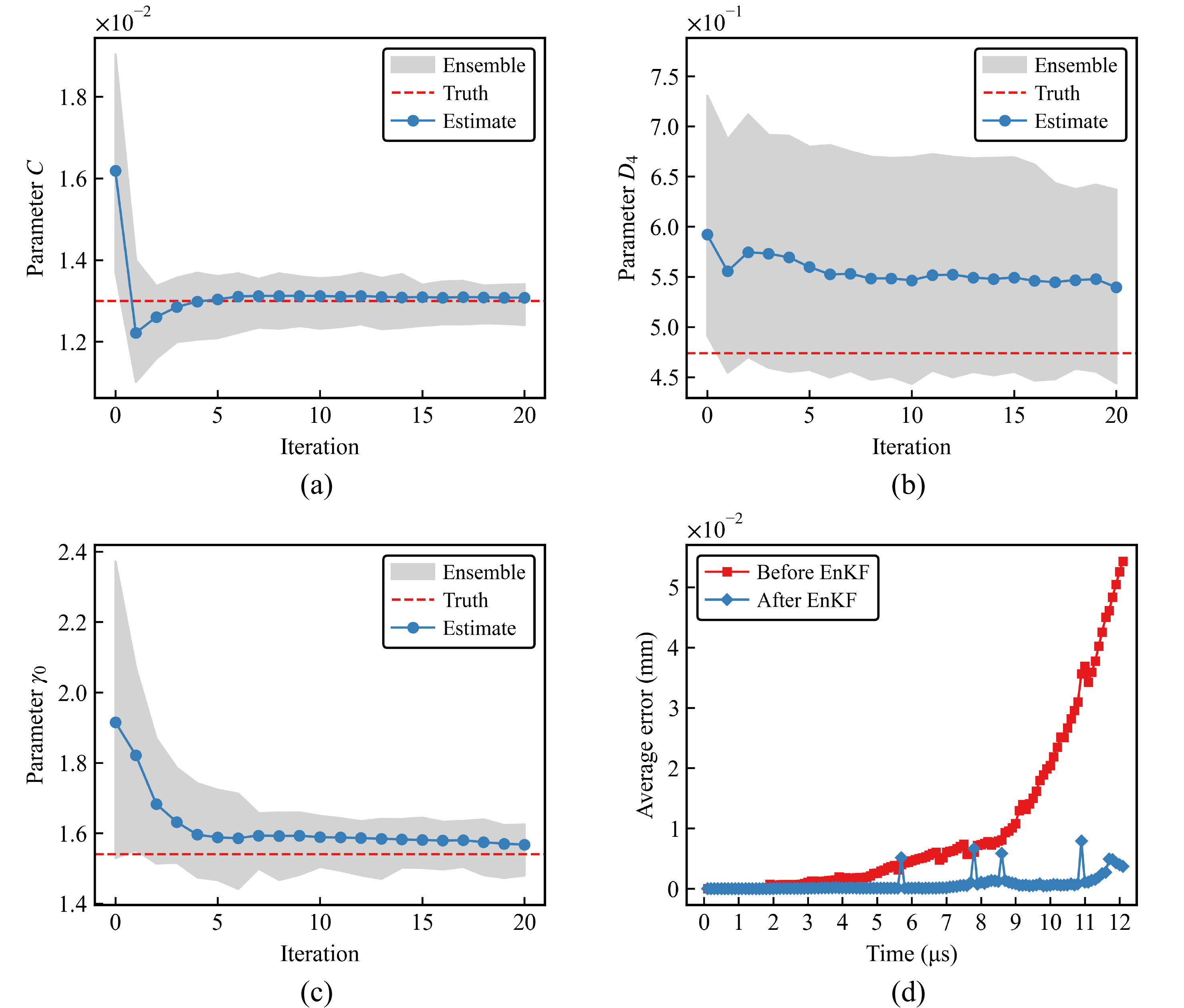}
\caption{EnKF results for Case 2 with initial ensemble mean $\bm{u}_0 = 1.25 \, \bm{u}_{\text{true}}$ ($+25\%$ bias): (a)-(c) Iteration histories of inferred parameters $C$, $D_4$, and $\gamma_0$. (d) Time-resolved RMSE of the back-face deflection before and after assimilation with respect to the true deflection over the $0$-$12 \, \upmu\rm{s}$ observation window.}
\label{fig:EnKF_result_2}
\end{figure}

\begin{figure}[!htbp]
\centering
\includegraphics[width=0.95\textwidth]{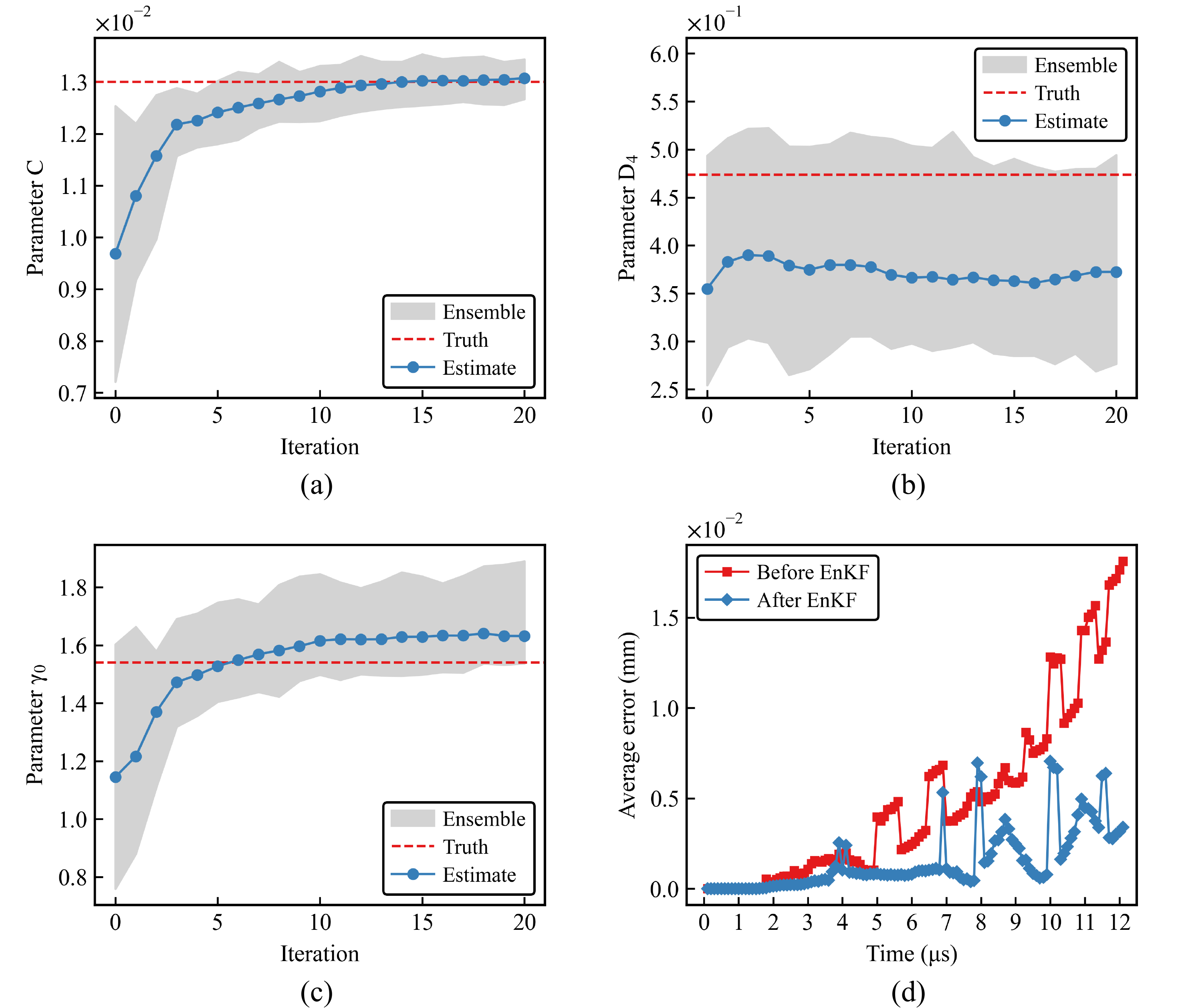}
\caption{EnKF results for Case 3 with initial ensemble mean $\bm{u}_0 = 0.75\,\bm{u}_{\text{true}}$, with $N_{\rm{o}}=10$ at each observation time: (a)-(c) Iteration histories of inferred parameters $C$, $D_4$, and $\gamma_0$. (d) Time-resolved RMSE of the back-face deflection before and after assimilation with respect to the true deflection over the $0$-$12 \, \upmu\rm{s}$ observation window.}
\label{fig:EnKF_result_3}
\end{figure}

\begin{figure}[!htbp]
\centering
\includegraphics[width=0.95\textwidth]{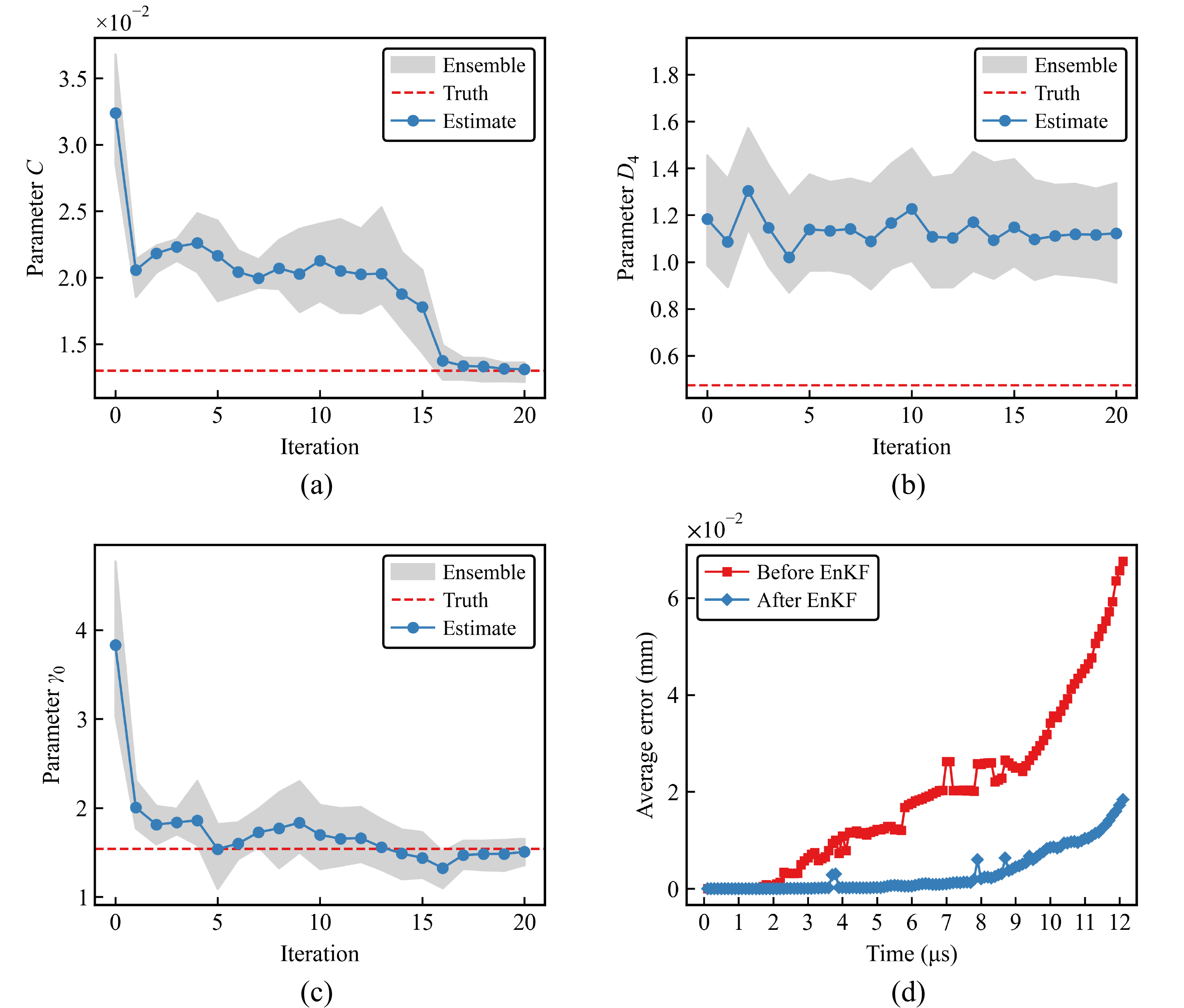}
\caption{EnKF results for Case 4 with initial ensemble mean $\bm{u}_0 = 2.5\,\bm{u}_{\text{true}}$ ($+150\%$ bias): (a)-(c) Iteration histories of inferred parameters $C$, $D_4$, and $\gamma_0$. (d) Time-resolved RMSE of the back-face deflection before and after assimilation with respect to the true deflection over the $0$-$12 \, \upmu\rm{s}$ observation window.}
\label{fig:EnKF_result_4}
\end{figure}

The EnKF results for the four cases are plotted in Figs.~\ref{fig:EnKF_result_1}-\ref{fig:EnKF_result_4}. All figures share a unified layout: subfigures~(a)-(c) display the iteration trajectories of $C$, $D_4$, and~$\gamma_0$, respectively, where the red dashed line marks the truth, the blue dots represent the ensemble mean, and the light-gray shading denotes the min-max ensemble range.
Subfigure~(d) reports the time-resolved RMSE of the simulated back-face deflection before and after assimilation, both computed against the reference displacement at the baseline values of the unknown parameters. In addition, a concise comparison of the ensemble mean, standard deviation (std), and relative error for all four cases is summarized in Table~\ref{tab:enkf_summary}.

In Case~$1$, Figs.~\ref{fig:EnKF_result_1}(a) and (c) show that EnKF efficiently corrects the $-25\%$ initial bias, with both $C$ and $\gamma_0$ converging almost monotonically to their true values within the first eight iterations. Table~\ref{tab:enkf_summary} confirms that the relative errors of $C$ and $\gamma_0$ decrease from $-25.50\%$ and $-25.62\%$ to $+0.91\%$ and $+1.92\%$, respectively. The posterior uncertainties are also greatly reduced. The standard deviation of $C$ drops from $\mathcal{O}(10^{-3})$ to $\mathcal{O}(10^{-4})$, and that of $\gamma_0$ decreases from $\mathcal{O}(10^{-1})$ to $\mathcal{O}(10^{-2})$. In contrast, the parameter $D_4$ remains poorly identified, as shown in Fig.~\ref{fig:EnKF_result_1}(b). Its ensemble mean does not converge to the true value, despite the true value being within the initial ensemble spread. The estimation error decreases from $-25.11\%$ to $-14.41\%$, while its posterior spread ($\sim 9.7\%$) is smaller than its prior spread ($\sim 13.1\%$). This outcome is consistent with the low sensitivity of $D_4$ observed in Fig.~\ref{fig:back-face_rmse}. 

Additionally, as shown in Fig.~\ref{fig:EnKF_result_1}(d), the RMSE of the back-face deflection decreases substantially, particularly during the late stage of the impact. For example, at $t = 12.0~\upmu\mathrm{s}$, it drops from about $2.0 \times 10^{-2}\,\mathrm{mm}$ to less than $1.0 \times 10^{-3}\,\mathrm{mm}$. Notably, although only the observations at three time instants, $t = 10.0~\upmu\mathrm{s}$, $11.0~\upmu\mathrm{s}$, and $12.0~\upmu\mathrm{s}$, are used to calibrate the material models, the post-calibration RMSE remains very low throughout the entire impact event (from $0$ to $12.0~\upmu\mathrm{s}$). This outcome demonstrates excellent predictive agreement with the synthetic observations once the sensitive parameters have been recovered. The intermittent spikes are manifestations of the forward solver, where numerical instabilities induce temporal discontinuities, resulting in sudden, non-physical variances in the predicted response.

In Case~$2$, Fig.~\ref{fig:EnKF_result_2} reveals a convergence behavior similar to that of Case~1, despite the $+25\%$ initial bias. The parameters $C$ and $\gamma_0$ again converge rapidly within the first five iterations, with their final errors reduced to $+0.63\%$ and $+1.76\%$, respectively. As summarized in Table~\ref{tab:enkf_summary}, the posterior standard deviation of $C$ and $\gamma_0$ decreases by roughly one order of magnitude. The parameter $D_4$ shows an initial downward trend over the first two iterations and briefly approaches the true value, but stabilizes after the tenth iteration with a persistent bias of approximately $7.5\%$. Fig.~\ref{fig:EnKF_result_1}(d) further demonstrates that the RMSE of the back-face displacement is markedly reduced after assimilation, decreasing by more than $90\%$ for $t \ge 9.0~\upmu\mathrm{s}$. Notably, whereas the pre-EnKF RMSE rises sharply after about $t = 9.0~\upmu\mathrm{s}$, the calibrated model exhibits only a slight increase over the same interval.

Both preceding cases use the same set of observations: $N_{\rm o}=20$ along an observation line of length $L=6.6$~mm from the center of the impact zone and exhibit excellent calibration precision. To evaluate the impact of the amount of observational data, we perform a further analysis of Case~$3$, wherein the number of observations is halved, specifically setting $N_{\rm o}=10$ and $L=3.3$~mm, while maintaining the initial estimate as in Case~$3$. The calibration results are presented in Fig.~\ref{fig:EnKF_result_3}. 
In circumstances where observational data are limited, the sensitive parameters $C$ and $\gamma_0$ exhibit the ability to convergence, although additional iterations are required to reach convergence. This is accompanied by final errors of $+0.56\%$ and $+5.97\%$, respectively, as documented in Table~\ref{tab:enkf_summary}.
The weakly identifiable parameter $D_4$ shows a continuing divergence: its mean drifts from $-25.11\%$ to $-21.38\%$, and its ensemble spread remains broad at approximately $9.1\%$ of the mean. Consistently, the RMSE of the back-face deflection after assimilation, shown in Fig.~\ref{fig:EnKF_result_3}(d), remains nearly identical to the results in Case~$1$ before $t \approx 6~\upmu\mathrm{s}$, however, exhibit slightly larger fluctuations and a higher terminal level after $t \approx 10~\upmu\mathrm{s}$.
These results underscore the importance of using full-field back-face measurements (for example, those obtained via 3D-DIC) to achieve reliable material-model calibration in HVI problems.

To stress-test the robustness of the EnKF inversion framework, we significantly bias the initial guess in Case~$4$ such that all the true parameters lie outside the initial ensemble spread. Fig.~\ref{fig:EnKF_result_4} summarizes the results. Even under this extreme misspecification, the filter remains numerically stable and effectively reduces predictive error and posterior uncertainty within the impact window. The two sensitive parameters, $C$ and $\gamma_0$, converge to their true values after approximately $15$ iterations. As reported in Table~\ref{tab:enkf_summary}, the initial discrepancies of about $+150\%$ for both parameters are reduced to $+0.81\%$ and $-2.16\%$, respectively, while their ensemble spreads decrease by roughly one order of magnitude. In contrast, the insensitive parameter $D_4$ remains biased ($+136.94\%$) and retains a relatively large ensemble spread, indicating that the posterior uncertainty appropriately reflects limited identifiability rather than exhibiting false confidence. Moreover, Fig.~\ref{fig:EnKF_result_4}(d) shows that the post-calibration RMSE decreases markedly, confirming that the calibrated model reproduces the observed response even under a severely biased initial guess.

It is worth noting that the proposed parameter rejuvenation strategy is not activated in Cases~$1$-$3$, because the sensitive parameters $C$ and $\gamma_0$ converge very quickly to their true values and the insensitive parameter $D_4$ retains a relatively large ensemble spread. However, this strategy plays a critical role in Case~$4$, where a strongly biased initial guess is used. In particular, Figs.~\ref{fig:EnKF_result_4}(a) and (c) show that parameter rejuvenation inflates the parameter covariance between iterations $5$ and $15$. Following this inflation, the parameter estimates migrate toward the true values. We further examine a baseline case with the same setup as Case~$4$ but without parameter rejuvenation, as detailed in Appendix~\ref{app:baseline}. In this baseline case, the ensemble variance collapses rapidly, causing the filter to lock into an incorrect solution before the mean can migrate toward the true parameter values.

\begin{table*}[!htbp]
\centering
\caption{Prior ($l=0$) and posterior ($l=20$) ensemble statistics.}
\label{tab:enkf_summary}

\sisetup{
  retain-explicit-plus,
  table-number-alignment = center,
  table-text-alignment   = center
}
\setlength{\tabcolsep}{2pt}
\footnotesize                
\resizebox{\linewidth}{!}{   
\begin{tabularx}{\linewidth}{
  l
  *{3}{>{\centering\arraybackslash}X 
        >{\centering\arraybackslash}S[table-format = +3.0]}}
\toprule\toprule
            & \multicolumn{2}{c}{\quad \quad \quad \quad $C$}
            & \multicolumn{2}{c}{\quad \quad \quad \quad $D_{4}$}
            & \multicolumn{2}{c}{\quad \quad \quad \quad $\gamma_{0}$}\\[-0.3em]
            & {mean $\pm$ std} & {error [\%]}
            & {mean $\pm$ std} & {error [\%]}
            & {mean $\pm$ std} & {error [\%]}\\
\midrule
\multicolumn{7}{@{}l}{\bfseries Case 1: under-biased initial guess, $\bm{u}_0=0.75\,\bm{u}_{\text{true}}$, $N_{\rm{o}}=20$}\\
\midrule
Prior      & $9.68\times10^{-3}\pm1.01\times10^{-3}$ & \num{-25.50}
           & $3.55\times10^{-1}\pm4.64\times10^{-2}$ & \num{-25.11}
           & $1.15\pm1.58\times10^{-1}$              & \num{-25.62}\\
Posterior  & $1.31\times10^{-2}\pm1.38\times10^{-4}$ & \num{+0.91}
           & $4.06\times10^{-1}\pm3.94\times10^{-2}$ & \num{-14.41}
           & $1.57\pm2.67\times10^{-2}$              & \num{+1.92}\\
\midrule
\multicolumn{7}{@{}l}{\bfseries Case 2: over-biased initial guess, $\bm{u}_0=1.25\,\bm{u}_{\text{true}}$, $N_{\rm{o}}=20$} \\
\midrule
Prior      & $1.62\times10^{-2}\pm1.00\times10^{-3}$ & \num{+25.00}
           & $5.92\times10^{-1}\pm4.64\times10^{-2}$ & \num{+24.95}
           & $1.92\pm1.58\times10^{-1}$              & \num{+24.38}\\
Posterior  & $1.31\times10^{-2}\pm1.52\times10^{-4}$ & \num{+0.63}
           & $5.39\times10^{-1}\pm4.01\times10^{-2}$ & \num{+13.86}
           & $1.57\pm2.37\times10^{-2}$              & \num{+1.76}\\
\midrule
\multicolumn{7}{@{}l}{\bfseries Case 3: limited observations, $\bm{u}_0=0.75\,\bm{u}_{\text{true}}$, $N_{\rm{o}}=10$}\\
\midrule
Prior      & $9.68\times10^{-3}\pm1.01\times10^{-3}$ & \num{-25.50}
           & $3.55\times10^{-1}\pm4.64\times10^{-2}$  & \num{-25.11}
           & $1.15\pm1.58\times10^{-1}$   & \num{-25.62}\\
Posterior  & $1.31\times10^{-2}\pm1.49\times10^{-4}$ & \num{+0.56}
           & $3.72\times10^{-1}\pm3.40\times10^{-2}$  & \num{-21.38}
           & $1.63\pm3.96\times10^{-2}$   & \num{+5.97}\\
\midrule
\multicolumn{7}{@{}l}{\bfseries Case 4: strongly biased initial guess,  $\bm{u}_0 = 2.5\,\bm{u}_{\text{true}}$, $N_{\rm{o}}=20$}\\
\midrule
Prior      & $3.24\times10^{-2}\pm1.26\times10^{-3}$ & \text{+149.5}
           & $1.18\pm9.10\times10^{-2}$  & \text{+149.89}
           & $3.84\pm3.25\times10^{-1}$   & \text{+149.38}\\
Posterior  & $1.31\times10^{-2}\pm2.00\times10^{-4}$ & +0.81
           & $1.12\pm8.28\times10^{-2}$  & +136.94
           & $1.51\pm4.46\times10^{-2}$   & -2.16\\
\bottomrule
\end{tabularx}}
\end{table*}

\begin{figure}[!htbp]
\centering
\includegraphics[width=0.95\textwidth]{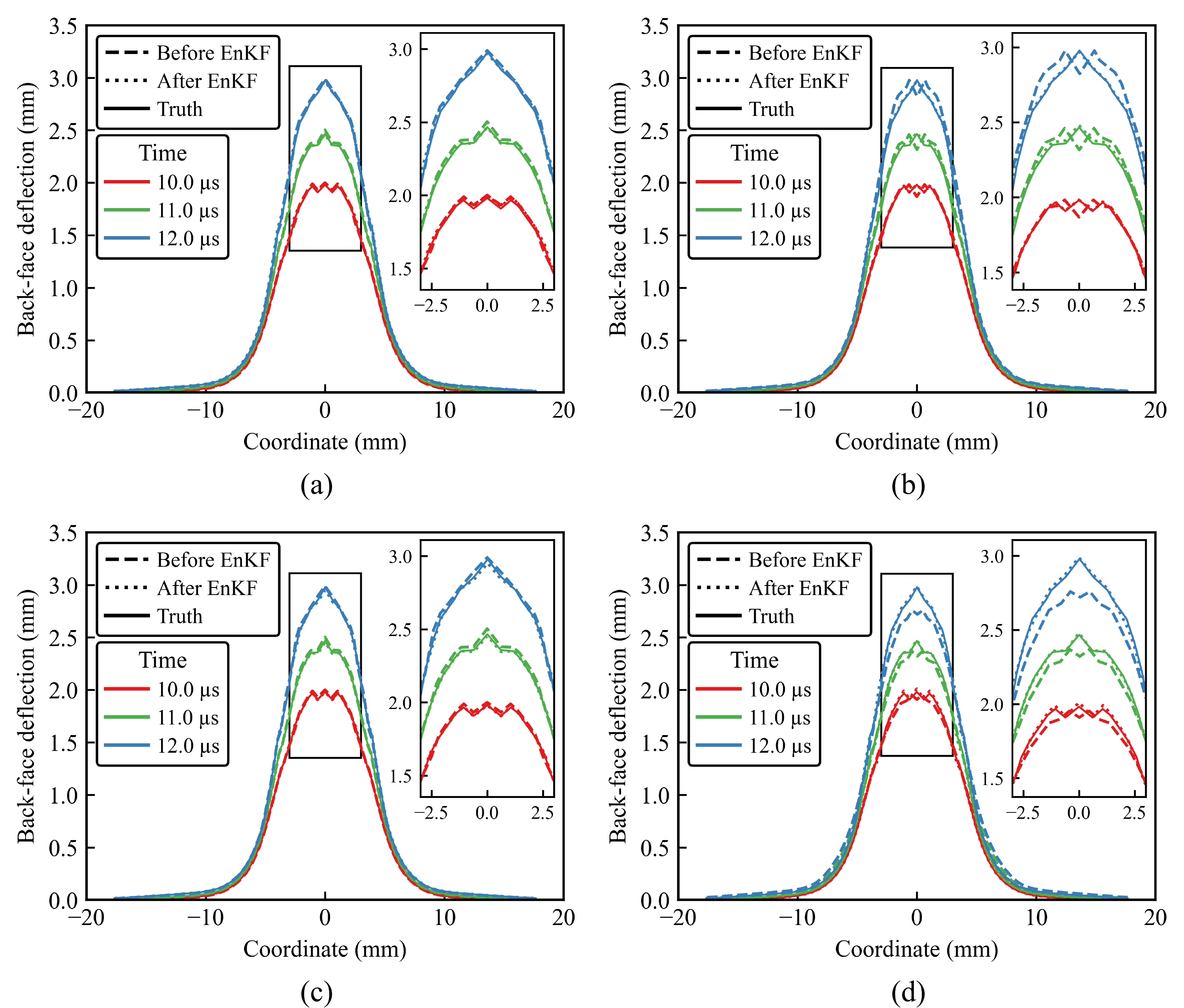}
\caption{Comparison of back-face deflection profiles at $t=10.0, 11.0$, and $12.0~\mu s$: (a) Case~1, (b) Case~2, (c) Case~3, and (b) Case~4. }
\label{fig:EnKF_comp}
\end{figure}

To visually assess the quality of the calibration, Fig.~\ref{fig:EnKF_comp} compares the back-face deflection profiles at the three representative time instants for Cases~1-4. Each figure overlays the baseline deflection computed at the true parameter values, the prediction from the initial guess (Prior), and the prediction from the calibrated model (EnKF Posterior). In Cases~2 and~4, where the prior prediction (dashed lines) shows a clear deviation from the baseline (solid lines), the calibrated profiles (dotted lines) closely align with the ground truth, demonstrating the effectiveness of the filter. In Cases~1 and~3, where the initial guess visually appears similar to the truth, the EnKF framework nevertheless accurately identifies the sensitive parameters $C$ and $\gamma_0$, providing calibrated results suitable for subsequent HVI simulations.

\begin{figure}[!htbp]
\centering
\includegraphics[width=0.9\textwidth]{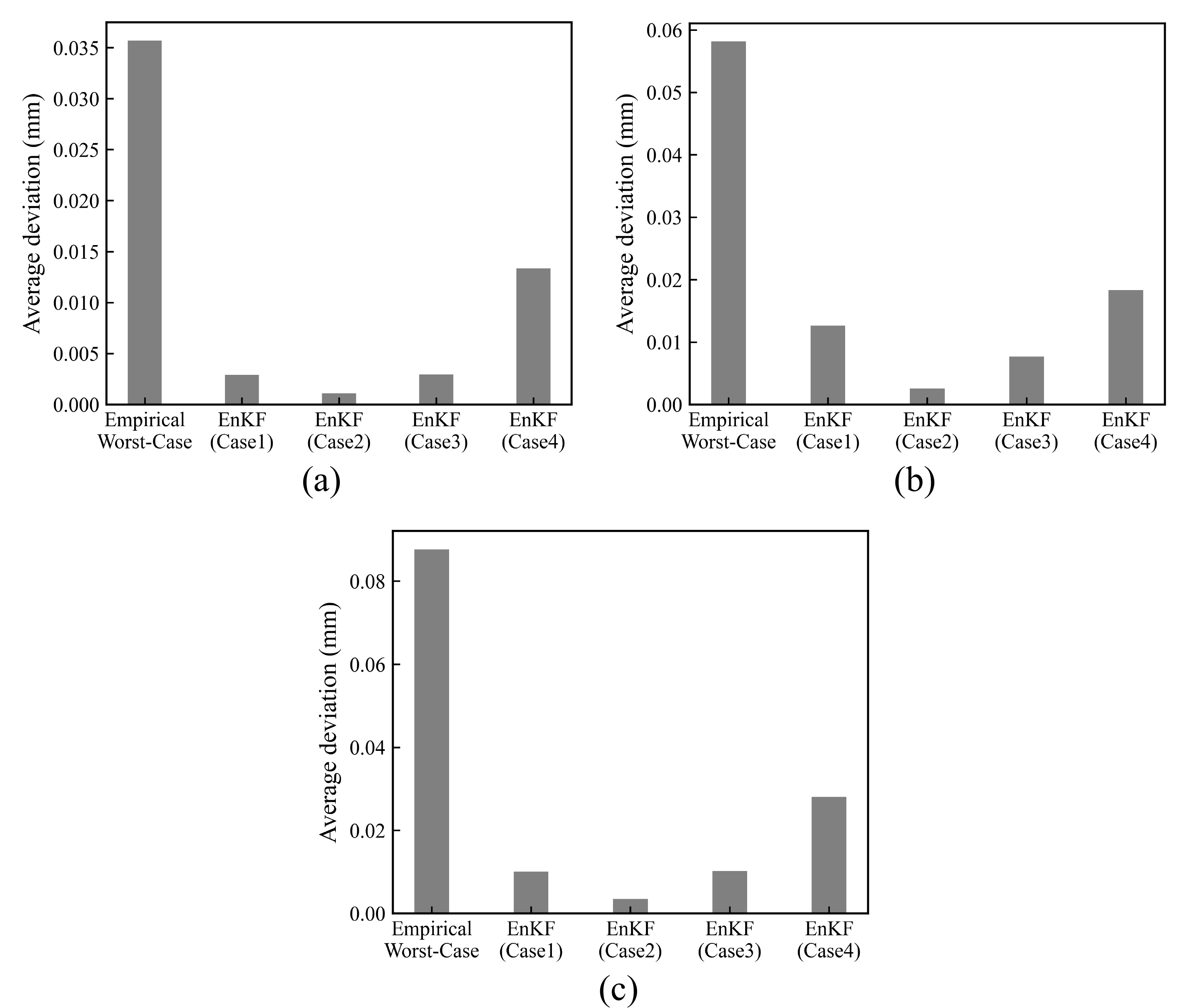}
\caption{Comparison of the post-calibration fitting error against the empirical sensitivity bound at: (a) $t=10.0~\upmu s$, (b) $t=11.0~\upmu s$, and (c) $t=12.0~\upmu s$. }
\label{fig:fit_error}
\end{figure}

To further verify that the low fitting error reflects accurate parameter recovery rather than model insensitivity, we benchmark the post-calibration RMSE against the overall parameter sensitivity, as shown in Fig.~\ref{fig:fit_error}. We perform a sampling-based screening to estimate an empirical sensitivity bound. Specifically, we evaluate the back-face deflections at the eight vertices of the $\pm 25\%$ parameter hypercube and at $20$ additional random samples uniformly distributed within these bounds. We use these back-face deflections to calculate the maximum deviation in the output induced by simultaneous variation of all three parameters. Notably, at all time instants, the post-calibration RMSE is substantially smaller than this empirical deviation. For example, the largest deviation among the sampled parameter sets is $8.77\times10^{-2}$~mm at $t=12.0~\upmu$s, whereas the EnKF-calibrated models achieve much lower errors. In Case~2, the RMSE at $t=12.0~\upmu$s is reduced to $3.51\times10^{-3}$~mm, roughly 25 times smaller than the empirical maximum. Even in Case~4, which starts from a strongly biased initial guess, the post-calibration RMSE is $2.80\times10^{-2}$~mm, still more than three times lower than the background variability. This order-of-magnitude reduction confirms that EnKF drives the data-model mismatch well below the inherent variability, demonstrating robust optimization of the dominant parameters ($C$ and $\gamma_0$) despite the weak identifiability of $D_4$.

\begin{figure}[!htbp]
\centering
\includegraphics[width=0.9\textwidth]{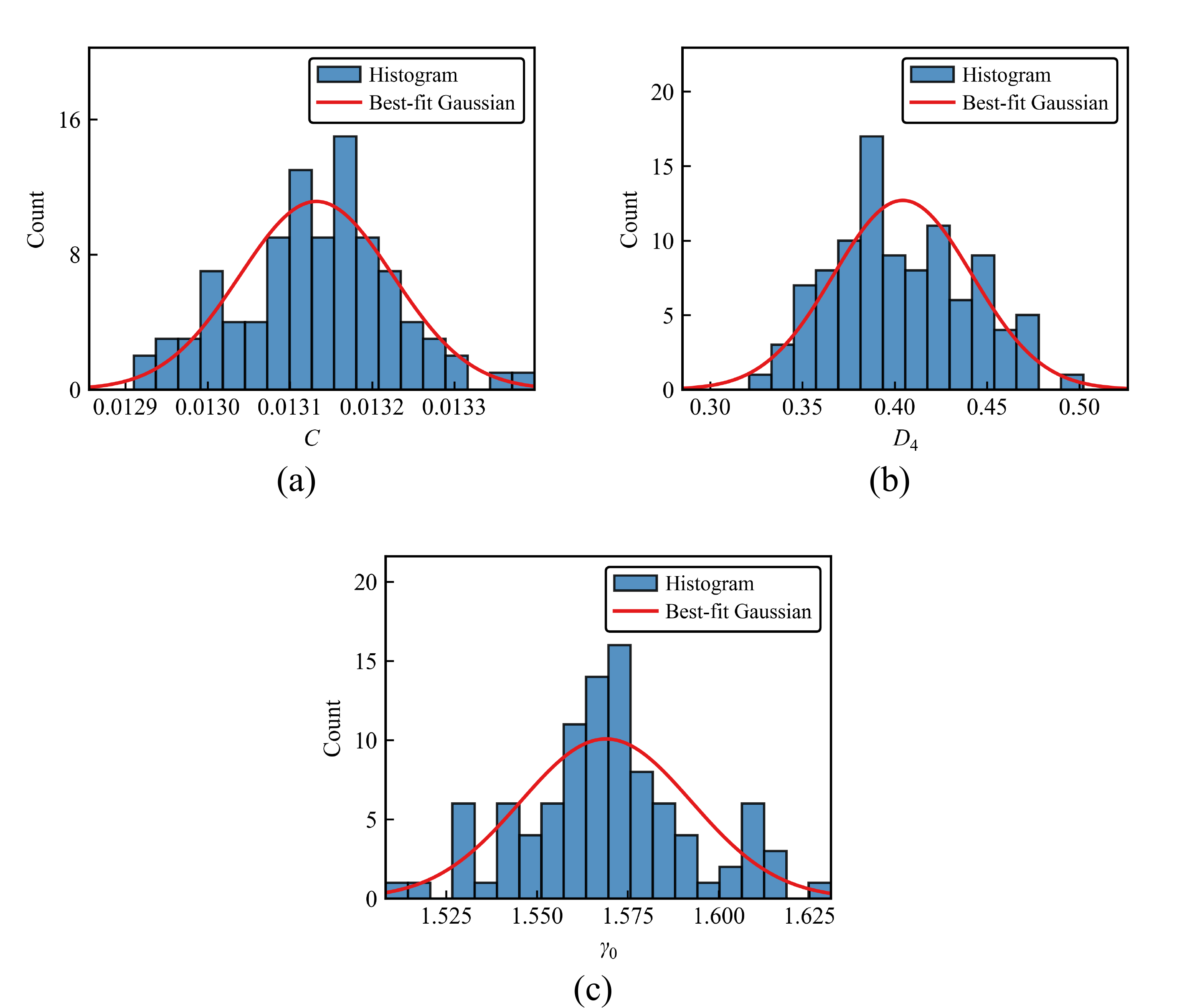}
\caption{Histogram of the final EnKF parameter estimates with fitted Gaussian curves for (a) $C$, (b) $D_4$, and (c) $\gamma_0$ in Case~$1$.}
\label{fig:EnKF_result_1_his}
\end{figure}

The estimation errors of the calibrated parameters can also be examined through the final EnKF ensemble. Fig.~\ref{fig:EnKF_result_1_his} shows histograms of the ensemble members with fitted Gaussian curves for the three parameters in Case~$1$. Because the EnKF framework propagates and updates only the first two statistical moments of the unknown parameters, it is important to verify that this assumption is valid. The near-Gaussian distributions observed in Fig.~\ref{fig:EnKF_result_1_his} confirm that these first two moments are sufficient for accurate calibration of the three material models, even within the highly nonlinear SPH simulation.

\section{Discussion} 
\label{sec:discussion}

Understanding and predicting the performance of materials under HVI play a critical role in both fundamental research and practical applications. However, experimental testing is often limited due to cost, time, and technical challenges. High-fidelity simulations provide a valuable alternative, but require accurate material models and reliable model parameters, including those for plasticity, fracture, and EOS. Traditionally, these parameters are estimated by manually tuning them until a visually acceptable agreement is achieved with the selected experimental results. For example, plasticity and fracture parameters can be obtained from Split-Hopkinson pressure and tension bar experiments, while EOS parameters are typically determined from plate impact tests. Nevertheless, such experiments are labor- and time-intensive, and the manual tuning process can introduce additional uncertainties. In this work, we propose an alternative approach based on EnKF: directly assimilating HVI observations into SPH simulations to characterize material properties. This strategy bypasses the need for separate calibration experiments and manual parameter tuning and enhances the predictive capability of simulations.

A key consideration for SPH-based inverse modeling is computational cost. Using a simple impact-inspired example, we have verified that EnKF is at least one order of magnitude more efficient than a traditional MCMC method in terms of the number of forward evaluations. This advantage can be further amplified by the inherent parallelism of EnKF. Within each EnKF iteration, the ensemble forward simulations are independent and can be executed concurrently, which is particularly beneficial for computationally expensive HVI simulations. In our numerical HVI studies, each SPH run takes approximately 10 minutes. If EnKF converges within $10$ iterations with an ensemble size of $100$, this corresponds to roughly $1{,}000$ forward runs in total. With $100$ parallel cores, this translates to about $100$ minutes of wall-clock time, or approximately $10$ hours using $16$ cores. In contrast, an RWM-MCMC implementation requiring $\mathcal{O}(10^{4})$ sequential steps (without surrogate models) would take on the order of $69$ days. These projections underscore that standard MCMC methods remain computationally prohibitive for this class of problems unless surrogate models are introduced, whereas EnKF offers a practical pathway for calibrating expensive black-box models by leveraging parallel computing to reduce wall-clock time.

We evaluate the proposed framework using synthetic back-face deflection data of the target plate. Experimentally, such data can be obtained through videography techniques, such as 3D-DIC. Beyond providing insights into material performance, these measurements can also serve as valuable input for calibrating numerical simulations. In this work, we verify the approach using a simplified scenario in which all measured data are assumed to have the same noise level. However, previous studies have shown that the noise in 3D-DIC often increases with the magnitude of displacement, as pixel-tracking uncertainty grows~{\cite{zhao2023uncertainty}}. This displacement-dependent noise can be readily incorporated into our framework.

We employ the EnKF-based framework to calibrate three models commonly used in HVI simulations: the JC plasticity model, the JC fracture model, and the Gr\"uneisen EOS. As discussed in Section~\ref{sec:method}, these models involve a total of $14$ parameters that must be determined. Directly calibrating all of these parameters within the EnKF framework can be challenging. To address this, we first perform an OAT sensitivity analysis prior to data assimilation. The analysis shows that certain parameters, such as $A$, $B$, and $n$ in the JC plasticity model and $D_1$, $D_2$, and $D_3$ in the JC fracture model, are highly sensitive to back-face deflection. However, these parameters are not included in the EnKF calibration because they can be independently measured through relatively simple experiments, such as conventional uniaxial tension tests. 

Conversely, parameters such as $D_4$ and $D_5$ in the JC fracture model are found to be insensitive to back-face deflection in selected observation time instances. We use $D_4$ as an example to demonstrate that insensitive parameters do not converge to their true values in the EnKF process, instead exhibiting large ensemble variance and hence significant uncertainty. This finding highlights an important consideration: in real applications where the true material parameters are unknown, both parameter sensitivity and the accuracy of calibration can be assessed by examining ensemble variance. If precise calibration of insensitive parameters is required, additional data must be incorporated, e.g. the back-face deflection at further time instances.

It is worth noting that the OAT sensitivity analysis serves only as a preliminary screening tool and does not fully capture parameter uncertainty across the input space. For a more comprehensive uncertainty quantification, methods such as McDiarmid’s subdiameters~\cite{sun2020rigorous, sun2022uncertainty, sun2023learning} can be employed, which measure the largest deviation in performance metrics resulting from finite variations in the corresponding input random variables.

Our studies also reveal a clear distinction between the effects of parameter insensitivity and limited data. As shown in Figs.~\ref{fig:EnKF_result_4}(a) and (c), although the EnKF estimates continue to converge towards the true values when the number of observations is diminished, they require additional iterations and exhibit larger ensemble standard deviations. This indicates that a convergence analysis on the quantity of available data is essential when using ensemble-based data assimilation techniques to calibrate material models. Sufficiently sensitive and abundant observational data are required to achieve accurate estimates and reduced uncertainty.

To further evaluate the scalability of the EnKF framework, we consider a higher-dimensional HVI problem parameterized by sensitive unknowns $\bm{u} = [A, C, D_1, D_2, C_s, \gamma_0]^{\text{T}}$. All other simulation settings are identical to those in Case~1, and detailed calibration results are reported in Appendix~\ref{app:enkf_case5}. The main finding is that $A$, $C$, $C_s$, and $\gamma_0$ converge to their true values, although they require more iterations than in Case~1. In contrast, $D_1$ and $D_2$ do not converge to their true values and retain relatively large ensemble spreads. Examination of the EnKF performance in the observation space suggests that this non-convergence and large uncertainty for $D_1$ and $D_2$ may arise from non-uniqueness with respect to the chosen observables: different combinations of these fracture parameters yield nearly indistinguishable back-face deflection profiles. Multiple realizations of the initial ensemble are tested, and this behavior persists consistently. This phenomenon underscores the ill-posed nature of HVI inverse problems in higher-dimensional parameter spaces, where distinct fracture mechanisms can compensate for one another to produce similar global responses.

This six-dimensional problem stands in sharp contrast to the three-parameter cases in the main text, where sensitive material parameters consistently converge to the true values with small uncertainties regardless of the initial prior bias. We emphasize again that OAT is used here only as a local sensitivity metric for pre-screening. Low RMSD values from the OAT analysis indicate low local sensitivity and non-uniqueness of the corresponding parameters. However, high RMSD values alone cannot guarantee uniqueness or identifiability. Therefore, in practical applications where the ground truth is unknown, a multi-start strategy---performing calibration from several distinct initial guesses---is essential to verify the robustness of convergence and to diagnose potential unidentifiability through persistently large ensemble variance.

\section{Concluding Remarks} 
\label{sec:summary}

In this study, we have developed an EnKF-based inversion framework to calibrate material models used in HVI simulations. The framework has been applied to identify representative JC plasticity, fracture, and Gr\"uneisen EOS parameters in SPH simulations of a steel ball impacting an AZ31B magnesium plate, using time-series measurements of back-face deflection. While other parameters are fixed based on literature or conventional tests following a sensitivity screening, the selected subset are used to demonstrate the framework's ability to recover sensitive parameters and diagnose non-identifiable ones. Subsequently, four three-dimensional and one six-dimensional cases have been designed to examine the influence of initial guesses, the quantity of observational data, and the number of unknowns on the calibration outcomes. From these investigations, the following conclusions can be drawn:
\begin{enumerate}
    \item Compared to traditional derivative-free Bayesian MCMC methods, the EnKF framework is at least one order of magnitude more efficient in terms of forward model evaluations. By leveraging parallel computing, the wall-clock time for calibration is reduced from months to hours, making it a practically feasible pathway for high-fidelity, black-box HVI simulations.
        
    \item With a sufficient amount of data, the EnKF framework efficiently recovers the sensitive model parameters, specifically $C$ in the JC plasticity model and $\gamma_{0}$ in the Gr\"uneisen EOS, within the first five iterations. The resulting estimates exhibit very low errors (mostly below $1\%$) and extremely small ensemble standard deviations (up to three orders of magnitude smaller than the mean). 

    \item The framework demonstrates robustness against extreme prior misspecification through a parameter rejuvenation strategy. Under $+150\%$ initial bias, this strategy prevents premature filter collapse, allowing sensitive parameters to migrate from outside the initial ensemble spread toward the true values. Without such intervention, the filter converges to incorrect estimates with small uncertainties.

    \item The high-dimensional inversion study highlights the fundamental challenge of equifinality in HVI calibration: multiple parameter combinations can produce virtually indistinguishable responses. It shows that, while the proposed framework delivers reliable predictive performance, parameter-level uniqueness in higher-dimensional settings can be compromised by strong correlations among influential parameters. Such non-uniqueness is reflected by persistently large ensemble uncertainties across EnKF iterations, and it may be alleviated by incorporating additional and more informative observations.
\end{enumerate}

Several future works can be investigated on the basis of the current conclusion. First, the proof-of-concept must be transitioned from synthetic to in-situ experimentation by replacing numerically generated back-face deflection fields with experimentally measured data, such as those obtained from high-speed 3D-DIC. In conjunction with this modification, it is imperative to integrate a heteroskedastic noise model, characterized by variance that amplifies with the extent of displacement, as observed in quantitative DIC uncertainty analysis~\cite{zhao2023uncertainty}, within the likelihood function to mitigate bias in the posterior distribution. Second, it is essential to extend the inverse framework beyond the JC or Gr\"uneisen hypothesis by integrating alternative constitutive models, which encompass a wider spectrum of projectile velocities, along with ancillary diagnostic techniques. These enhancements will enable the simultaneous calibration of fracture coefficients akin to those referenced in $D_4$, which are otherwise non-identifiable solely through deflection measurement data, and will support application to a diverse range of alloy systems. Ultimately, since the EnKF regards the forward solver as a black-box computational module, the present LS-DYNA SPH forward model may be replaced by surrogate and reduced-order models developed through machine learning regression, without necessitating alterations to the inversion framework. These advances are expected to provide improved predictive accuracy and significant reductions in computational expense, thus expanding the practical applicability of the proposed data assimilation methodology.

\section{Acknowledgement} 
R.J. and X.S. gratefully acknowledge the support of the National Science Foundation (NSF), USA under Grant No. $2429424$, and the support of the University of Kentucky, USA through the Faculty Start-up Fund. G.W. gratefully acknowledges the support of the National Natural Science Foundation of China (52301336), Science and Technology Development Fund of Macau S.A.R. (0048/2025/ITP1), and University of Macau (SRG2025-00004-FST). R.J. and X.S. appreciate the insightful discussions with Dr. K.~T. Ramesh, Dr. Suhas Eswarappa Prameela, Dr. Pinkesh Malhotra, and Dr. Justin Moreno. The suggestions of the anonymous reviewers have helped to improve the quality and scope of this work. 
\appendix
\section*{Appendix}
\section{Baseline EnKF Performance under Extreme Bias}
\label{app:baseline}

\begin{figure}[htbp!]
    \centering
    \includegraphics[width=1\textwidth]{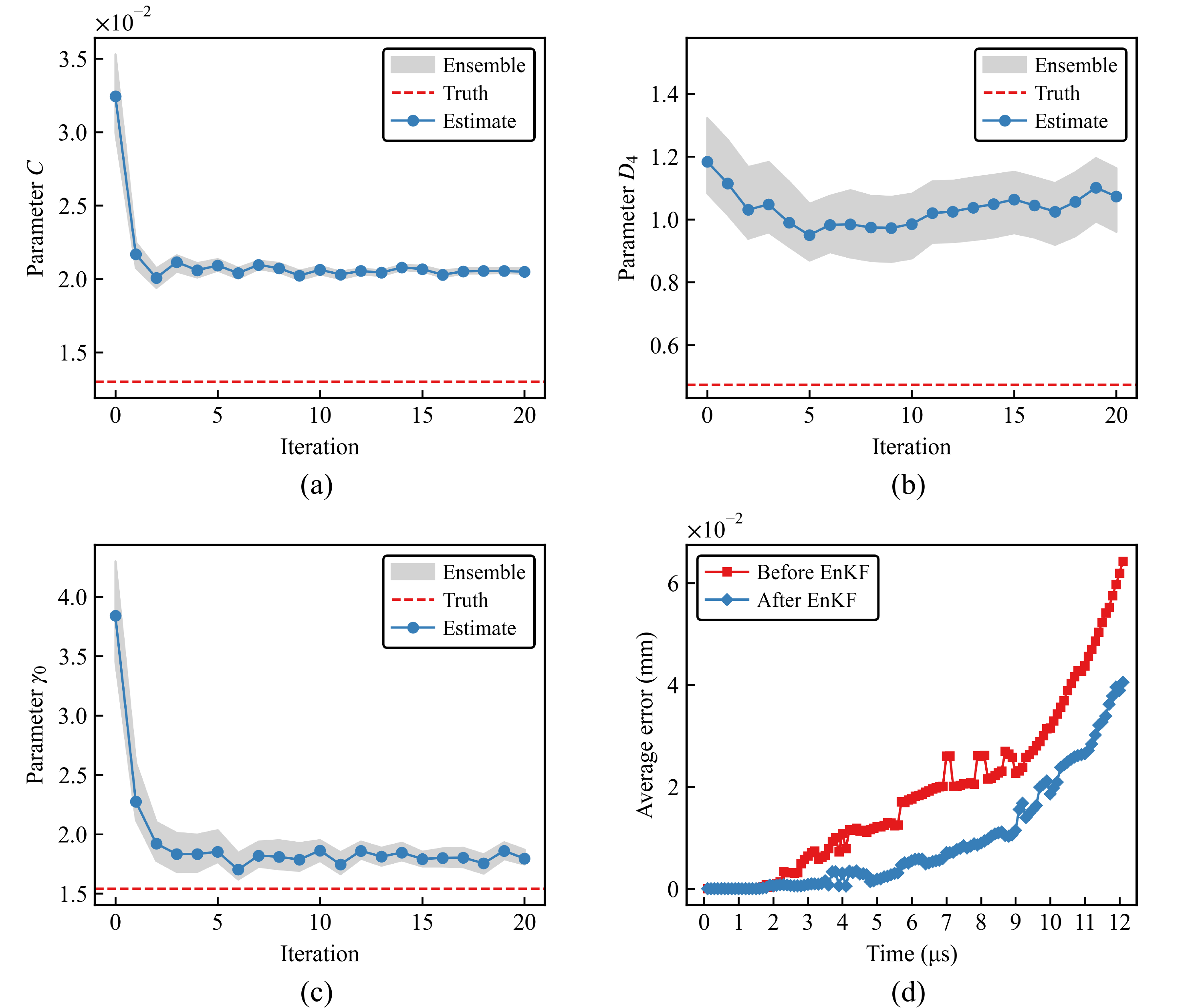} 
    \caption{EnKF results for Case 4 without parameter rejuvenation: (a)-(c) Iteration histories of inferred parameters $C$, $D_4$, and $\gamma_0$. (d) Time-resolved RMSE of the back-face deflection before and after assimilation with respect to the true deflection over the $0$-$12 \, \upmu\rm{s}$ observation window.}
    \label{fig:EnKF_case_4o}
\end{figure}

To demonstrate the necessity of the rejuvenation strategy introduced in Section~\ref{subsec:CI}, we present here the results of Case 4 using the EnKF implementation only with RTPS inflation under the same extreme initialization ($\bm{u}_0 \approx 2.5\,\bm{u}_{\text{true}}$). As illustrated in Fig.~\ref{fig:EnKF_case_4o}, the standard filter exhibits a characteristic ``drift-then-stall'' behavior. The ensemble variance collapses rapidly within the first 3-5 iterations (indicated by the narrowing grey band), causing the filter to lock into an incorrect solution before the mean can migrate to the true values. Quantitatively, while the initial discrepancies were approximately $+150\%$ for all parameters, the standard EnKF could only reduce the errors of the sensitive parameters $C$ and $\gamma_0$ to $+57.63\%$ and $+16.50\%$, respectively. This contrasts starkly with the results of Case 4 using the triggered rejuvenation method, which achieved errors of $+0.81\%$ and $-2.16\%$. Furthermore, the standard filter reports a deceptively small posterior standard deviation for these biased estimates, a phenomenon known as ``false confidence''. This comparison highlights that without parameter rejuvenation, the filter is unable to recover from strong prior biases even when parameters are identifiable.

\section{EnKF Inversion For a High-Dimensional Case}
\label{app:enkf_case5}

\begin{figure}[!htbp]
\centering
\includegraphics[width=0.95\textwidth]{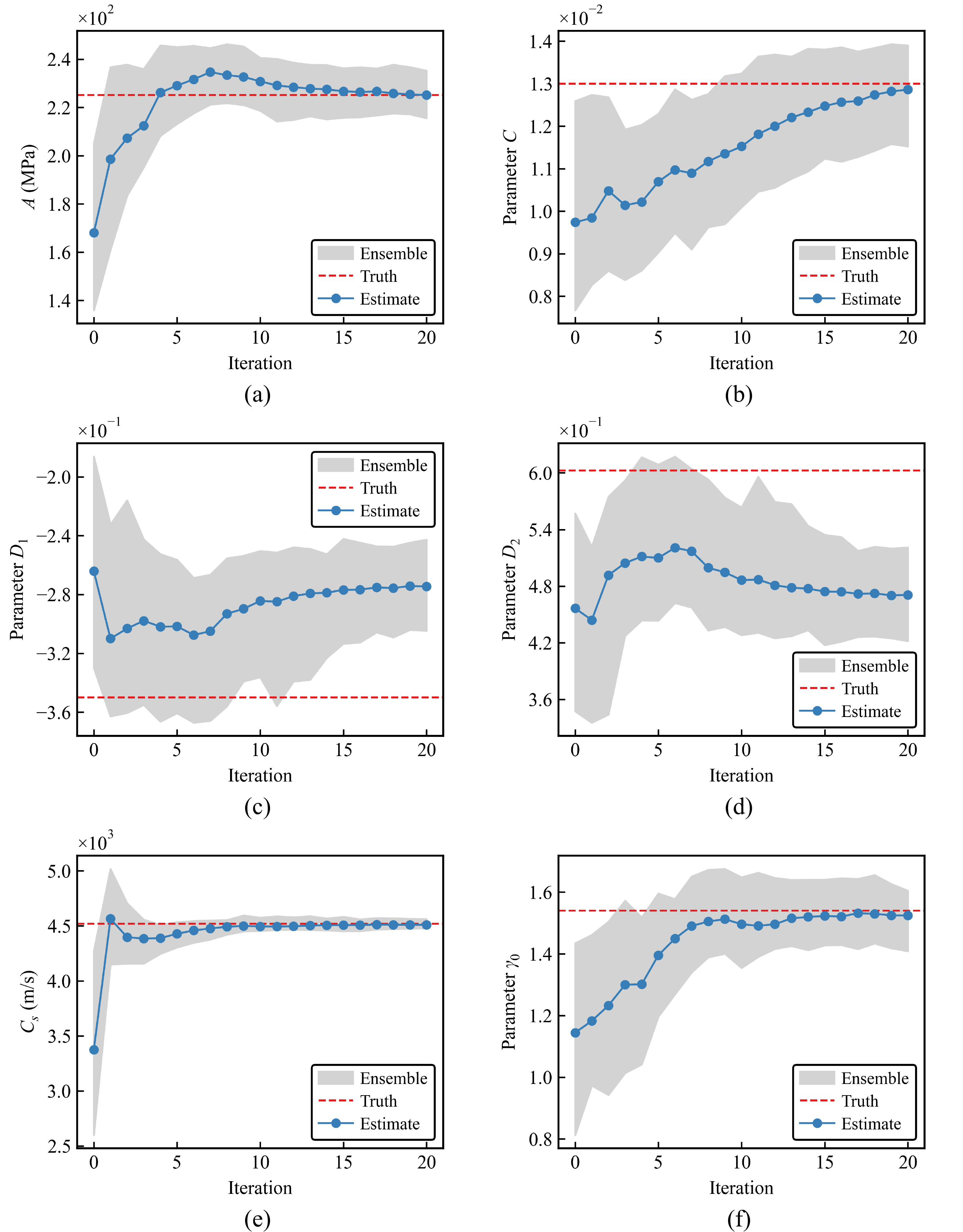}
\caption{EnKF inversion results for the high-dimensional case with 6 unknown parameters. (a)-(f) Iteration histories of the ensemble mean (blue line) and spread (gray area) for parameters $A, C, D_1, D_2, C_s, \gamma_0$. The red dashed line indicates the true value.}
\label{fig:EnKF_result_5}
\end{figure}

We introduce a high-dimensional problem: an expanded parameter vector defined as $\bm{u} = [A, C, \allowbreak D_1, D_2, \allowbreak C_s, \gamma_0]^\text{T}$. In this scenario, the calibration scope is expanded to six parameters, selecting the most sensitive ones from each material model: $A$ and $C$ (plasticity), $D_1$ and $D_2$ (fracture), and $C_s$ and $\gamma_0$ (EOS). The initial ensemble is generated with a $-25\%$ bias (${\bm u}_0 = 0.75 \bm u_{\rm true}$), consistent with the setup in Case 1, and the same observation set ($N_o=20$) is utilized. The inversion results for this case are presented in Fig.~\ref{fig:EnKF_result_5} and summarized in Table~\ref{tab:enkf_case5}.

The parameters $A$, $C$, $C_s$, and $\gamma_0$ converge toward their true values, albeit requiring more iterations than in Case~1 as expected. Table~\ref{tab:enkf_case5} shows that the relative errors for $A$ and $C_s$ are reduced to $+0.07\%$ and $-0.23\%$, respectively. $C$ and $\gamma_0$ achieve final errors of approximately $1\%$. The posterior uncertainties for these four parameters are also substantially reduced, indicating strong identifiability. In contrast, the fracture parameters $D_1$ and $D_2$ do not converge to their true values. Although they were identified as sensitive in the OAT analysis in Section~\ref{subsec:MIO}, they collapse to incorrect values and retain relatively large ensemble spreads, with residual biases of $-21.6\%$ and $-21.9\%$, respectively. This outcome highlights that OAT, as a local sensitivity metric, does not guarantee identifiability in a higher-dimensional inverse problem.

\begin{figure}[!htbp]
\centering
\includegraphics[width=0.95\textwidth]{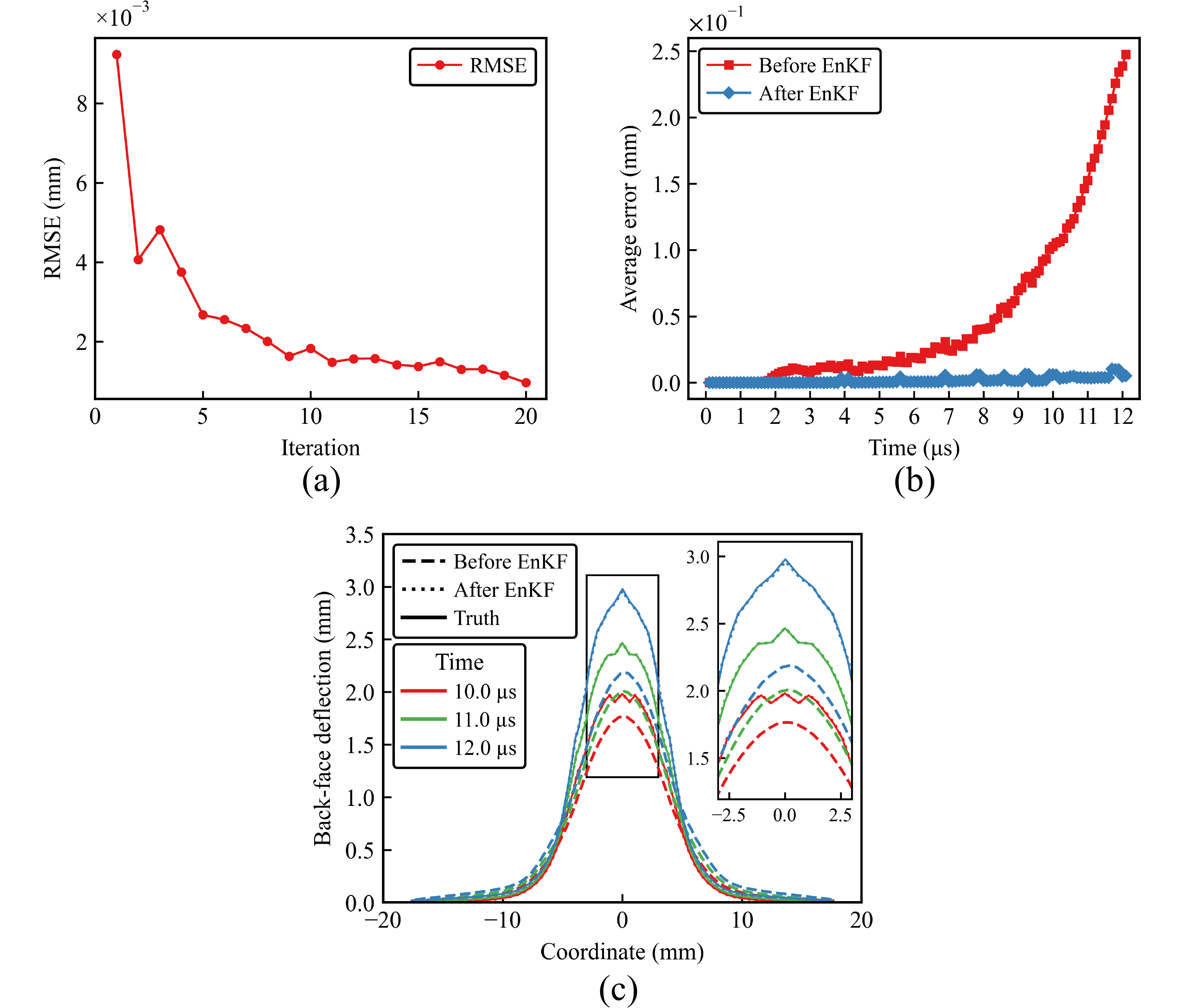}
\caption{Assessment of predictive capability for the high-dimensional case in the observation space: (a) Iteration history of the observation RMSE. (b) Time-resolved average error of the back-face deflection before ($l=0$) and after ($l=20$) assimilation. (c) Comparison of back-face deflection profiles at $t=10.0, 11.0$, and $12.0~\mu s$.}
\label{fig:EnKF_case5_rmse}
\end{figure}

To understand why $D_1$ and $D_2$ do not converge to their true values, we examine the performance of the EnKF framework in the observation space, as shown in Fig.~\ref{fig:EnKF_case5_rmse}. Despite clear deviations in parameter space, the calibrated model achieves excellent predictive accuracy in the observable. Specifically, Fig.~\ref{fig:EnKF_case5_rmse}(a) shows that the RMSE of the observation vector decreases rapidly and stabilizes at a negligible level ($\sim 10^{-4}$~mm) after about ten iterations. Fig.~\ref{fig:EnKF_case5_rmse}(b) further confirms that the time-resolved error in the back-face deflection remains consistently small throughout the impact process. The posterior back-face deflection profiles in Fig.~\ref{fig:EnKF_case5_rmse}(c) are in excellent agreement with the true response, effectively correcting the initial bias at the output level.

Taken together, Figs.~\ref{fig:EnKF_result_5} and~\ref{fig:EnKF_case5_rmse} may provide an evidence of non-uniqueness with respect to the chosen observable. They show that different parameter sets---namely, the true set versus the calibrated set with biased $D_1$ and $D_2$---can produce virtually indistinguishable back-face deflection profiles. This indicates that, in the present regime, the available back-face deflection data, while sufficient to constrain stiffness and compressibility (dominated by $A, C, C_s, \gamma_0$), remain compatible with multiple combinations of the fracture initiation and evolution parameters ($D_1, D_2$). This finding underscores an important limitation: although the EnKF framework reliably ensures predictive accuracy for the observable quantity, the physical interpretability of individual parameters hinges on identifiability, and strict uniqueness may require additional diagnostic data (e.g., perforation hole diameter, residual velocity, or back-face at more time instances) to break such parameter correlations.

\begin{table}[htbp]
\centering
\caption{Summary of EnKF results for the high-dimensional case.}
\label{tab:enkf_case5}
\small
\renewcommand{\arraystretch}{1.2}
\begin{tabular}{lcccc}
\hline
Parameter & Truth & Prior mean $\pm$ std & Posterior mean $\pm$ std & Error [\%] \\
\hline
$A$ [MPa]   & $2.25\times10^2$  & $1.68\times10^2 \pm 1.32\times10^1$ & $2.25\times10^{2} \pm 3.58$ & $+0.07$ \\
$C$     & $1.30\times10^{-2}$   & $9.74\times10^{-3} \pm 9.55\times10^{-4}$ & $1.29\times10^{-2} \pm 4.98\times10^{-4}$ & $-1.06$ \\
$D_1$   & $-3.50\times10^{-1}$  & $-2.64\times10^{-1} \pm 2.69\times10^{-2}$ & $-2.75\times10^{-1} \pm 1.17\times10^{-2}$ & $-21.60$ \\
$D_2$   & $6.03\times10^{-1}$   & $4.57\times10^{-1} \pm 4.58\times10^{-2}$ & $4.71\times10^{-1} \pm 1.91\times10^{-2}$ & $-21.90$ \\
$C_s$ [m/s] & $4.520\times10^3$ & $3.38\times10^{3} \pm 3.37\times10^{2}$ & $4.51\times10^{3} \pm 1.53 \times10^{1}$ & $-0.23$ \\
$\gamma_0$  & $1.54$            & $1.14 \pm 1.21\times10^{-1}$ & $1.52 \pm 4.03\times10^{-2}$ & $-1.01$ \\
\hline
\end{tabular}
\end{table}

\bibliographystyle{unsrtnat}
\color{black}
\bibliography{references_}

\end{document}